\documentclass[11pt]{article}
\usepackage{latexsym}
\usepackage{graphicx}
\usepackage{caption}
\usepackage{psfrag}
\usepackage{amsmath,amssymb,dsfont}
\newcommand{\Frac}[2]{\frac{\displaystyle #1}{\displaystyle #2}}
\newcommand{\beq}{\begin{equation}}
\newcommand{\eeq}{\end{equation}}
\newcommand{\beqn}{\begin{eqnarray}}
\newcommand{\eeqn}{\end{eqnarray}}
\newcommand{\beqns}{\begin{eqnarray*}}
\newcommand{\eeqns}{\end{eqnarray*}}

\input epsf

\numberwithin{equation}{section}

\begin{document}

\begin{titlepage}

\begin{flushright}
LMU-ASC 82/11\\
USTC-ICTS-11-15
\end{flushright}
\vspace*{1.5cm}
\begin{center}
{\Large \bf $K^+\to\pi^+\pi^0e^+e^-:$ a novel short-distance probe}\\[3.0cm]

{\bf L. Cappiello}$^{a,b}$, {\bf O. Cat\`{a}}$^{c,d}$, {\bf G. D'Ambrosio}$^{b}$ and {\bf Dao-Neng Gao}$^{e}$\\[1cm]

$^{a}${\it{\small{Dipartimento di Scienze Fisiche,
             Universit\'a di Napoli "Federico II", Via Cintia, 80126 Napoli, Italy}}}\\[.5 cm]
$^{b}${\it{\small{INFN-Sezione di Napoli, Via Cintia, 80126 Napoli, Italy}}}\\[.5 cm]
$^{c}${\it{\small{Departament de F\'isica Te\`orica and IFIC, Universitat de Val\`encia-CSIC,
Apartat de Correus 22085, E-46071 Val\`encia, Spain}}}\\[.5 cm]
$^{d}${\it{\small{Ludwig-Maximilians-Universit\"at M\"unchen, Fakult\"at f\"ur Physik, Arnold Sommerfeld Center for Theoretical Physics, D–80333 M\"unchen, Germany}}}\\[.5 cm]
$^{e}${\it{\small{Interdisciplinary Center for Theoretical Study and Department of Modern Physics,
University of Science and Technology of China, Hefei, Anhui 230026 China}}}

\end{center}

\vspace*{1.0cm}

\begin{abstract}
We study the decay $K^+\to\pi^+\pi^0e^+e^-$, currently under analysis by the NA62 Collaboration at CERN. In particular, we provide a detailed analysis of the Dalitz plot for the long-distance, $\gamma^*$-mediated, contributions (Bremsstrahlung, direct emission and its interference). We also examine a set of asymmetries to isolate genuine short-distance effects. While we show that charge asymmetries are not required to test short distances, they provide the best environment for its detection. This constitutes by itself a strong motivation for NA62 to study $K^-$ decays in the future. We therefore provide a detailed study of different charge asymmetries and the corresponding estimated signals.  Whenever possible, we make contact with the related processes $K^+\to\pi^+\pi^0\gamma$ and $K_L\to\pi^+\pi^-e^+e^-$ and discuss the advantages of $K^+\to\pi^+\pi^0e^+e^-$ over them.
\end{abstract}

\end{titlepage}

\setcounter{footnote}{0}


\section{Introduction}

Kaon physics has played a crucial role in establishing the CKM flavor structure of the Standard Model (see, for instance,~\cite{D'Ambrosio:1996nm,Barker:2000gd,Sozzi:2003ve,Cirigliano:2011ny}). In the recent years, the experimental status of rare kaon decays has improved to the point that, at the time of writing, the NA62 physics program at CERN is aiming to detect eighty $K^+ \to \pi ^+ \nu  \bar{\nu}$ events~\cite{NA62}. This constitues an extremely important test of the Standard Model (SM) and, as a result, a tool to constrain new physics~\cite{Smith:2010dr}. The NA62 experiment aims to produce a total of $10^{13}$ $K^+$'s, and hopefully maybe even $10^{14}$ in the future. While many $K^+$ decays have been studied in the past, it is timely that studies be devoted to $K^+$ decays which have deserved less attention but that, given the experimental situation, can provide important information on CP violation, new physics and chiral tests. 
    
Traditionally, kaon decays have been divided in three categories: i) short-distance physics dominated decays, like the aforementioned $K^+\to\pi^+\nu\bar{\nu}$~\cite{Artamonov:2008qb,PDG}; ii) decays where short and long-distance contributions are comparable, like $K_{L}\to\pi^0e^+e^-$~\cite{Buchalla:2003sj}; and iii) long-distance dominated decays, like radiative kaon decays~\cite{D'Ambrosio:1992bf,D'Ambrosio:1994ae}. However, even in this latter case, a particular charge, CP or angular asymmetry makes it possible to disentangle a small but still interesting short-distance contribution~\cite{Smith:2010dr,Mertens:2011ts}. Some examples of this last category are the CP charge asymmetries studied by NA48/2 in $K^\pm\to\pi^{\pm}\pi^0\pi^0$~\cite{:2007yfa} and $K^\pm\to\pi^\pm\pi^0\gamma$~\cite{:2010uja}.  

In particular, $K^\pm\to\pi^\pm\pi^0\gamma$ requires a very accurate Dalitz plot analysis to disentangle the Bremsstrahlung contribution, accurately predicted from QED, from the direct (electric and magnetic) emission amplitudes~\cite{D'Ambrosio:1994du,D'Ambrosio:2000yc,Cappiello:2007rs}. This Dalitz plot analysis is also useful to extract, directly from the NA62 measurement of the $K^\pm\to\pi^\pm\pi^0\gamma$  charge asymmetry, the interference term proportional to the CP violating component of the electric emission piece, leading to a bound for the CP asymmetry, namely $ A^{\pi^\pm\pi^0\gamma}_{CP}<1.5 \cdot 10^{-3}$~\cite{:2010uja}.  
  
In this paper we will study the decay $K^+\to\pi^+\pi^0e^+e^-$. This decay also belongs to category iii) above and is dominated by one-photon exchange, {\it{i.e.}}, $K^+\to\pi^+\pi^0\gamma^*\to\pi^+\pi^0e^+e^-$. Its amplitude can be parametrized as 
\begin{align}
{\cal M}_{LD}&=J_{em}^{\mu}(k_+,k_-)\frac{1}{q^2}H_\mu(p_1,p_2,q),
\end{align} 
where $p_i$ are the pion momenta, $k_i$ the lepton momenta and $q^2$ is the invariant mass of the dilepton pair. $H_{\mu}$ is the hadronic vector, which, based on Lorentz and gauge invariance, can be written in terms of 3 form factors $F_{1,2,3}$ as
\begin{equation}
H^\mu(p_1,p_2,q)=F_1 p^\mu_1+F_2 p_2^\mu+F_3 \varepsilon^{\mu\nu\alpha\beta}p_{1\nu}p_{2\alpha}q_\beta, 
\end{equation}
where $F_{1,2}$ are dominated by the Bremsstrahlung (B) contribution, while $F_3$ contains the Magnetic (M) piece. These different contributions were studied in considerable detail in Ref.~\cite{Pichl:2000ab}. In particular, one finds the following rate for the branching ratios:  
\begin{align}\label{ratioPi} 
 B(K^+\to\pi ^+ \pi^0 e ^+  e^-)  &\sim&   B(K^+ \to   \pi ^+ \pi^0 e ^+  e^-)_{B}  &\,\,\,\,\,\,+&  B(K^+ \to\pi ^+ \pi^0 e ^+  e^-)_M\nonumber\\
&=& (330\pm 15)\cdot 10^{-8}   &\,\,\,\,\,\,+&  ( 6.14\pm 1.3)\cdot 10^{-8}.  
\end{align}
These numbers clarify how hard one must work to extract dynamical information from this decay: much like with $K^{\pm}\to\pi^{\pm}\pi^0\gamma$, one needs to fight the dominant Bremsstrahlung contribution. The numbers in Eq.~(\ref{ratioPi}) were found by numerically integrating over the phase space using arbitrary kinematical variables~\cite{Pichl:2000ab}. One of the conclusions of the present work is that an improvement can be obtained by studying the Dalitz plot with a specific set of kinematical variables, namely $T_c^*$ and $E_{\gamma}^*$ (the kinetic energies of the charged pion and the photon, respectively, in the kaon rest frame) and the dilepton invariant mass $q^2$. $E_{\gamma}^*$ and $T_c^*$ were already introduced in Ref.~\cite{Christ:1967zz} to study $K^+\to\pi^+\pi^0\gamma$ and thus adding $q^2$ comes as a natural extension. We will show that the set $(E_{\gamma}^*,T_c^*,q^2)$ allows to identify the most dense regions of the Dalitz plot for each dynamical contribution. 

Quite generically, as $q^2$ increases the relative weight of the Bremsstrahlung contribution gets reduced. Additionally, and most interestingly, for increasing $q^2$ the different contributions tend to fill different regions of the $(E_{\gamma}^*,T_c^*)$ Dalitz plot. Thus, kinematical cuts in $(E_{\gamma}^*,T_c^*)$ can help discriminate the different dynamical contributions. Obviously, at high-$q^2$ the phase space closes and thus experimental detection is afflicted with poor statistics. However, we will show that there exists a rather wide optimal window around $q^2\sim (50$ MeV$)^2$. 

The ability to gauge the values of $q^2$ is an evident advantage of $K^+\to\pi^+\pi^0 e^+e^-$ over $K^+\to\pi^+\pi^0\gamma$, where one is restricted to $q^2=0$, and thus exposed to a large Bremsstrahlung background. Even so, the Dalitz plot analysis of $K^{\pm} \to\pi^{\pm} \pi^0\gamma$ has already been a useful tool to extract electric interference and direct emission amplitudes~\cite{:2010uja}. Besides improving on the accuracy of those determinations, $K^+\to\pi^+\pi^0 e^+e^-$ has a novel interference between Bremsstrahlung and direct magnetic emission. This provides a determination of the full $F_3$, including the strong phases associated with final state interactions. 

In this article, we will also investigate in detail short-distance physics in this channel. Similarly to $K_L\to\pi^+\pi^-e^+e^-$\cite{Sehgal:1992wm,Heiliger:1993qt,Elwood:1995xv,Elwood:1995dj}, short-distance effects can be obtained in $K^+\to\pi^+\pi^0e^+e^-$ by looking into P-violation in the lepton pair. As a result, and opposed to $K^{\pm}\to\pi^{\pm}\pi^0\gamma$, charge asymmetries are not needed for that purpose. However, in sharp contrast to $K_L\to\pi^+\pi^-e^+e^-$, in $K^{\pm}\to\pi^{\pm}\pi^0e^+e^-$ there is no $\epsilon-$type contribution and, consequently, every CP-violating signal in the charged decay is a genuine short-distance effect. 

We will be interested in two different short-distance aspects, namely i) CP violation induced by $K^+\to\pi^+\pi^0\gamma^*$ and ii) short distances associated with effective dimension-6 operators. For the latter, we will consider the contribution of SM physics as well as physics beyond the SM. In this last case, and in order to have a more predictive scenario, we will assume SM completions with Minimal Flavor Violation (MFV)~\cite{D'Ambrosio:2002ex}. We will show that, once MFV is assumed, rather precise statements can be made about short-distance signal detection.  

This paper is organized as follows: in Sect.~\ref{sec:II} we will start with an analysis of the kinematics. Sect.~\ref{sec:III} is devoted to the dynamics of $K^+\to\pi^+\pi^0\gamma^*$, which will then be used in Sect.~\ref{sec:IV} for the study of the long-distance contributions to $K^+\to \pi^+\pi^0 e^+e^-$. In particular, we will concentrate on the Dalitz plot of the photon-mediated P-conserving contribution in the ($E_{\gamma}^{\ast},T_c^{\ast}$) plane for different cuts in $q^2$. In Sect.~\ref{sec:V} we will turn to the study of short distances with the evaluation of the SM contribution. Sect.~\ref{sec:VI} is devoted to the study of P and CP-violating signals in this decay within and beyond the SM. Conclusions are given in Sect.~\ref{sec:VII}. 


\section{Kinematical preliminaries}\label{sec:II}

In this Section we discuss the kinematics of $K^+\to\pi^+\pi^0e^+e^-$ in terms of the dynamical variables $(E_{\gamma}^*,T_c^*,q^2,\theta_{\ell},\phi)$, to be defined below. Since we are interested in a full Dalitz plot analysis, we will begin by providing the analytic expression for its phase space. 

The invariant phase space for the four-body decay is defined as 
\begin{equation}
d\Phi=(2\pi)^4\delta^{(4)}(P-p_1-p_2-k_+-k_-)\frac{d^3p_1}{(2\pi)^32
E_1}\frac{d^3p_2}{(2\pi)^32 E_2}\frac{d^3k_+}{(2\pi)^32
E_+}\frac{d^3k_-}{(2\pi)^32 E_-},
\end{equation}
which can be trivially rewritten as
\begin{align}
d\Phi &=\int d^4 p_{\pi} \int d^4 q (2\pi)^4
\delta^{(4)}(P-p_{\pi}-q)\frac{d^3p_1}{(2\pi)^32
E_1}\frac{d^3p_2}{(2\pi)^32 E_2}
\delta^{(4)}(p_1+p_2-p_{\pi})\nonumber\\
&\times~ \frac{d^3k_+}{(2\pi)^32 E_+}\frac{d^3k_-}{(2\pi)^32
E_-}\delta^{(4)}(k_++k_--q),
\end{align}
where $p_{\pi}=p_1+p_2$ and $q=k_++k_-$ are the momenta of the dipion and dilepton pairs, respectively. Then one can obtain
\begin{align}
d\Phi &=\frac{1}{4m_K^2}(2\pi)^5\int d s_\pi \int d
s_\ell \lambda^{1/2}(m_K^2,p_{\pi}^2,q^2)\Phi_{\pi}\Phi_{\ell},
\end{align}  
where 
\begin{align}
\Phi_{\pi}&=
\int \frac{d^3p_1}{(2\pi)^32
E_1}\frac{d^3p_2}{(2\pi)^32 E_2} \delta^{(4)}(p_1+p_2-p_{\pi})=\frac{1}{(2\pi)^5}\frac{1}{8p_{\pi}^2}\lambda^{1/2}(p_{\pi}^2,m_{\pi^+}^2,m_{\pi^0}^2)\int
d\cos\theta_\pi,\nonumber\\
\Phi_{\ell}&= \int
\frac{d^3k_+}{(2\pi)^32
E_+}\frac{d^3k_-}{(2\pi)^32 E_-} \delta^{(4)}(k_++k_--q)=\frac{1}{(2\pi)^6}\frac{1}{8}\sqrt{1-\frac{4m_\ell^2}{q^2}}\int
d\phi\int d\cos\theta_\ell,
\end{align}
with $\lambda(a,b,c)=a^2+b^2+c^2-ab-ac-bc$. The angles above are defined as in Ref.~\cite{Cabibbo:1965zz}: if ${\bf{p_1}}$ is the $\pi^+$ momentum in the dipion CM system; ${\bf{k_+}}$ the $e^+$ momentum in the dilepton CM system; ${\bf{{\hat{n}}}}$ the direction of the dipion system as seen from the $K^+$ rest frame; and ${\bf{p_1^{\perp}}}$ and ${\bf{k_+^{\perp}}}$ the components of ${\bf{p_1}}$ and ${\bf{k_+}}$ perpendicular to ${\bf{{\hat{n}}}}$, then 
\begin{equation}
\cos\theta_{\pi}=\frac{\mathbf{\hat{n}}\cdot\mathbf{p_1}}{|\mathbf{p_1}|};\,\,\,\,\, \cos\theta_{\ell}=-\frac{\mathbf{\hat{n}}\cdot\mathbf{k_+}}{|\mathbf{k_+}|};\,\,\,\,\,
\cos\phi=\frac{\mathbf{p_1^{\perp}}\cdot\mathbf{k_+^{\perp}}}{|\mathbf{p_1^{\perp}}||\mathbf{k_+^{\perp}}|}.
\end{equation}
Intuitively, $\theta_{\ell}$ is the angle between the $e^+$ momentum and the dipion system as measured from the dilepton CM while $\phi$ is the angle between the dipion and dilepton planes.

The final result for the phase space therefore reads
\begin{equation}\label{ps31}
d^5\Phi=\frac{1}{2^{14}\pi^6m_K^2}\frac{1}{s_\pi}\sqrt{1-\frac{4m_\ell^2}{q^2}}
\lambda^{1/2}(m_K^2,p_{\pi}^2,q^2)\lambda^{1/2}(p_{\pi}^2,m_{\pi^+}^2,m_{\pi^0}^2)d
p_{\pi}^2 d q^2 d\cos\theta_\pi d\cos\theta_\ell d\phi,
\end{equation}
where the range of the kinematical variables is
\begin{align}\label{Kinevar}
4m_{\ell}^2 \, &\le \, q^2\, \le \,(m_K-\sqrt{p_{\pi}^2})^2,\nonumber\\
(m_{\pi^+}+m_{\pi^0})^2 \, &\le \, p_{\pi}^2 \, \le \,(m_K-2
m_\ell)^2,\nonumber\\
0\,&\le \, \theta_\pi,~\theta_\ell \, \le \, \pi,\nonumber\\
0 \, &\le \, \phi \, \le \, 2\pi.
\end{align}

In order to disentangle most easily the different contributions, it is convenient to work with the set of variables $(E_{\gamma}^{\ast},T_c^{\ast},q^2,\theta_{\ell},\phi)$, where $T_c^*$ and $E_{\gamma}^*$ are the kinetic energy of the charged pion and the photon, respectively, in the $K^+$ rest frame, and $q^2$ is the dilepton invariant mass. This set of variables is the natural extension of the variables introduced in Ref.~\cite{Christ:1967zz} to study $K^{\pm}\to \pi^{\pm}\pi^0\gamma$ decays. We can easily trade $p_{\pi}^2$ and $\cos\theta_{\pi}$ for $E_{\gamma}^*$ and $T_c^*$ using the relations:
\begin{eqnarray}\label{anglepion}
p_{\pi}^2&=&m_K^2-2m_KE_{\gamma}^*+q^2,\nonumber\\
\cos\theta_\pi&=&\frac{2(u-T_c^*)}{\beta \sqrt{{E_\gamma^*}^2-q^2}},
\end{eqnarray}
where 
\begin{align}
\beta=\frac{1}{p_{\pi}^{2}}\lambda^{1/2}(p_{\pi}^2,m_{\pi^+}^2,m_{\pi^0}^2), \quad\,\, u=\displaystyle\frac{m_K-E^*_\gamma}{2}\zeta_+-m_{\pi^+},\quad\,\, \zeta_{\pm}=\displaystyle\left(1\pm\frac{\delta m^2}{p_{\pi}^2}\right),
\end{align}
and $\delta m^2=m_{\pi^+}^2-m_{\pi^0}^2$ is an isospin breaking parameter. In terms of the variables $(E_{\gamma}^*,T_c^*,q^2,\theta_{\ell},\phi)$ the phase space takes the simple form
\begin{equation}\label{ps3}
d^5\Phi=\frac{1}{2^{11} \pi^6}\sqrt{1-\frac{4m_\ell^2}{q^2}}d E_\gamma^*
d q^2 d T_c^* d \cos\theta_\ell d \phi,
\end{equation}
and the limits of integration of Eq.~(\ref{Kinevar}), in terms of the set of variables $(E_{\gamma}^*,T_c^*,q^2,\theta_{\ell},\phi)$, become
\begin{align}\label{limits}
u-\frac{\beta}{2} \sqrt{{E_\gamma^*}^2-q^2}\,&\le \, T_c^* \, \le \,
u+\frac{\beta}{2} \sqrt{{E_\gamma^*}^2-q^2},\nonumber\\
\sqrt{q^2}\, &\le \, E^*_\gamma \, \le \, \frac{m_K^2-(m_{\pi^+}+m_{\pi^0})^2+q^2}{2m_K},\nonumber\\
2m_\ell \, &\le \, \sqrt{q^2} \, \le \, m_K-(m_{\pi^+}+m_{\pi^0}),\nonumber\\
0 \, &\le \, \theta_{\ell}\, \le \, \pi,\nonumber\\
0 \, &\le \, \phi \,\le \, 2\pi. 
\end{align}
For later convenience we also list the following expressions:
\begin{align}\label{conversion}
q\cdot p_{(1,2)}&=\frac{1}{2}(m_KE_\gamma^*-q^2)\zeta_{\pm}\mp m_K (u-T_c^*),\nonumber\\
p_1\cdot p_2&=\frac{1}{2}(m_K^2-2m_KE_{\gamma}^*+q^2-(m_{\pi}^2+m_{\pi^0}^2)),\nonumber\\
\epsilon_{\mu\nu\lambda\rho}p_1^{\mu}p_2^{\nu}q^{\lambda}Q^{\rho}&=\frac{m_K\zeta_-\beta_{\ell}}{2}\sqrt{q^2[m_K(m_K-2E_{\gamma}^*)+q^2][\beta^2(E_{\gamma}^{*2}-q^2)-4(u-T_c^*)^2]}\sin{\theta_{\ell}}\sin\phi,\nonumber\\
Q\cdot p_{(1,2)}&=\cos\theta_{\ell}\beta_{\ell}\left[-\frac{\beta[m_K(m_K-2E_{\gamma}^*)+q^2]}{4}\zeta_{\pm}\pm\frac{(m_KE_{\gamma}^*-q^2)(u-T_c^*)}{4\sqrt{E_{\gamma}^{*2}-q^2}}\right]\nonumber\\
&\mp\frac{\beta_\ell\sin\theta_{\ell}\cos\phi}{2}\sqrt{\frac{q^2[m_K(m_K-2E_{\gamma}^*)+q^2][\beta^2(E_{\gamma}^{*2}-q^2)-4(u-T_c^*)^2]}{E_{\gamma}^{*2}-q^2}},
\end{align}
which are needed to express the form factors and the kinematical weights (see next Section) in terms of $(E_{\gamma}^*,T_c^*,q^2,\theta_{\ell},\phi)$.


\section{Dynamics}\label{sec:III}
The amplitude for $K^+(P)\to \pi^+(p_1)\pi^0(p_2) \gamma^*(q)\to \pi^+(p_1)\pi^0(p_2) e^+(k_+)
e^-(k_-)$ can be parametrized as 
\begin{align}\label{amplitude1}
{\cal M}_{LD}=\frac{e}{q^2} \big[\bar{u}(k_-)\gamma^\mu v(k_+)\big] H_{\mu}(p_1,p_2,q),
\end{align} 
where $H_{\mu}$ is the hadronic vector, which can be written in terms of 3 form factors $F_{1,2,3}$:
\begin{equation}\label{param}
H^{\mu}(p_1,p_2,q)=F_1 p^\mu_1+F_2 p_2^\mu+F_3 \varepsilon^{\mu\nu\alpha\beta}p_{1\nu}p_{2\alpha}q_\beta.
\end{equation}
In terms of the momenta $p_{1,2}$, $q=k_++k_-$ and $Q=k_+-k_-$, the squared amplitude can be cast as
\begin{align}\label{squared}
\sum_{\rm spins}|{\cal M}_{LD}|^2&=\frac{2e^2}{q^4}\Bigg[\sum_i^3|F_i|^2T_{ii}+2{\mathrm{Re}}\sum_{i<j}^3(F_i^{*}F_j)T_{ij}\Bigg],
\end{align}
where
\begin{align}\label{Tkinematic}
T_{ij}&=q\cdot p_i q\cdot p_j-Q\cdot p_i Q\cdot p_j-q^2
p_i\cdot p_j;\qquad \qquad \qquad \qquad \,\,\,\,\, (i,j=1,2),\nonumber\\
T_{i3}&=(Q\cdot p_i)\epsilon_{\mu\nu\lambda\rho}p_1^{\mu}p_2^{\nu}q^{\lambda}Q^{\rho};\qquad \qquad \qquad\qquad\qquad\qquad\qquad\,\,\,\,(i=1,2),\nonumber\\
T_{33}&=4 m_\ell^2\big[(m_{\pi}^4-(p_1\cdot p_2)^2) q^2-m_{\pi}^2((q\cdot p_2)^2+(q\cdot
p_1)^2)+2p_1\cdot p_2q\cdot p_1 q\cdot p_2\big]\nonumber\\
&+(Q\cdot p_1)^2\big[(q\cdot p_2)^2-q^2m_{\pi}^2\big]+(Q\cdot p_2)^2\big[(q\cdot p_1)^2-q^2m_{\pi}^2\big]\nonumber\\
&+2Q\cdot p_1Q\cdot p_2(q^2 p_1\cdot p_2-q\cdot p_1 q\cdot p_2).
\end{align}
\begin{figure}[t]
\begin{center}
\includegraphics[width=5.5cm]{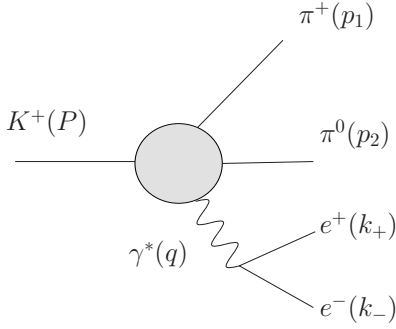}
\end{center}
\caption{\small{\it{Photon-mediated $K^+\to\pi^+\pi^0e^+e^-$ decay with our kinematic conventions. The blob represents the hadronic tensor $H_{\mu}$.}}}
\end{figure} 
\begin{figure}[t]
\begin{center}
\includegraphics[width=10cm]{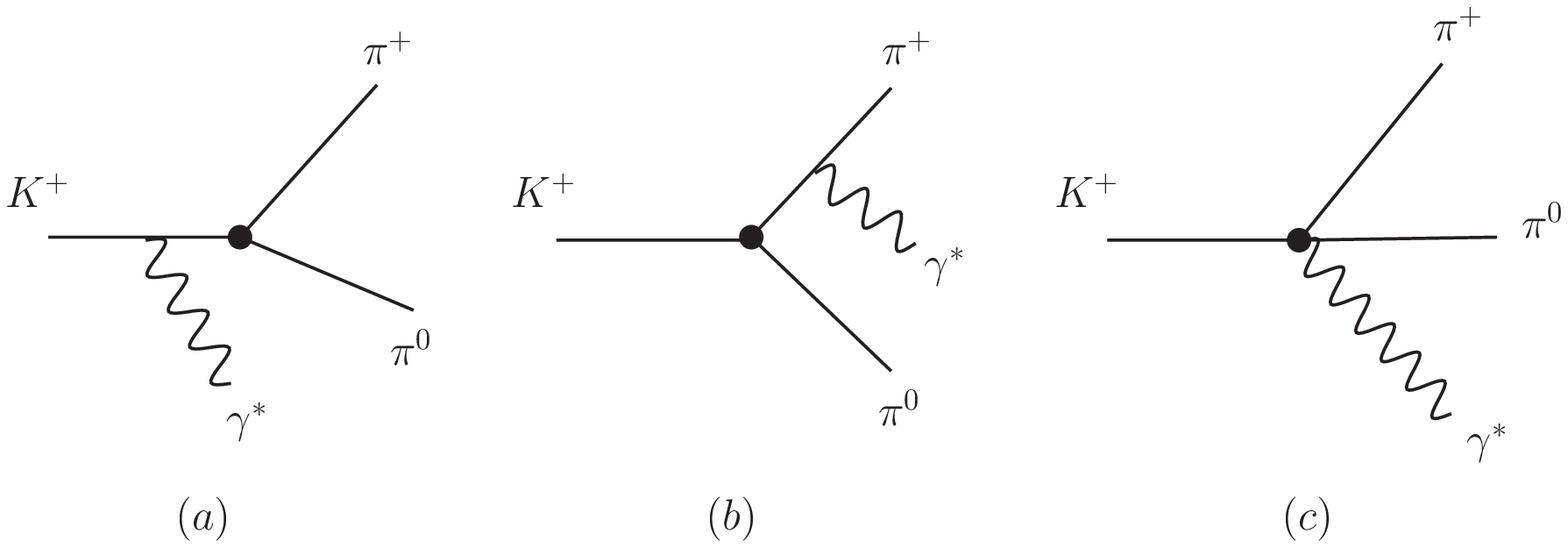}
\hskip 2.6cm
\includegraphics[width=3.1cm]{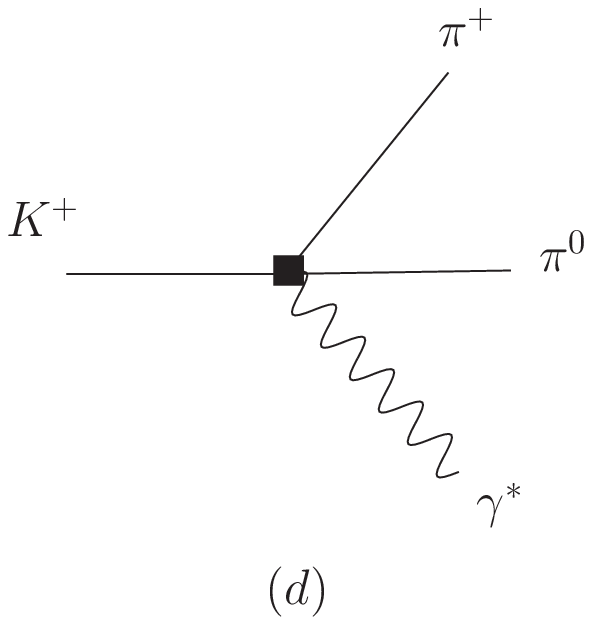}
\end{center}
\caption{\small{\it{Diagrams corresponding to the Bremsstrahlung (upper row) and direct emission (lower row) contributions. Dotted and squared vertices represent ${\cal{O}}(p^2)$ and ${\cal{O}}(p^4)$ weak operator insertions, respectively. Gauge invariance of the Bremsstrahlung contribution is ensured by Low's theorem.}}}
\end{figure} 
For phenomenological purposes, it is useful to split the form factors in terms of the (dominant) Bremsstrahlung and the direct emission contributions as $F_i=F_i^{(B)}+F_i^{(DE)}$, where $i=1,2$. 

According to Low's theorem~\cite{Low:1958sn}, the Bremsstrahlung piece can be written as
\begin{eqnarray}\label{brem}
{\cal{M}}(K^+\to \pi^+\pi^0\gamma^*)_{B}&=&2e\left[\frac{P\cdot \epsilon}{(P-q)^2-m_{K}^2}+\frac{p_1\cdot \epsilon}{(p_1+q)^2-m_{\pi^+}^2}\right]{\cal{M}}(K^+\to \pi^+\pi^0)\nonumber\\
&=&e\left[-\frac{P^{\mu}}{(P\cdot q)-\frac{q^2}{2}}+\frac{p_1^{\mu}}{(p_1\cdot q)+\frac{q^2}{2}}\right]\epsilon_{\mu}{\cal{M}}(K^+\to \pi^+\pi^0),
\end{eqnarray} 
from which one immediately concludes that 
\begin{align}\label{ib} 
F_1^{(B)}&=\frac{4ie(q\cdot p_2)}{(2 q\cdot
p_1+q^2)(2 q\cdot P-q^2)}{\cal{M}}(K^+\to\pi^+\pi^0),\nonumber\\
F_2^{(B)}&=\frac{-2ie}{2 q\cdot P-q^2} {\cal{M}}(K^+\to\pi^+\pi^0).
\end{align}
The expressions for ${\cal{M}}(K^+\to\pi^+\pi^0)$ and the direct emission piece (electric and magnetic) can be systematically computed using Chiral Perturbation Theory (ChPT)~\cite{Pichl:2000ab}. To leading order and neglecting isospin-violating effects they read,\footnote{Knowing that $BR(K^+\to \pi^+\pi^0)=0.2066(8)$ and $\Gamma_t=5.32\cdot 10^{-14}$ MeV~\cite{PDG}, one can easily work out that $|{\cal{M}}(K^+\to\pi^+\pi^0)|^2=3.28\cdot 10^{-16}$GeV$^{2}$. We will use this value in our analysis.}
\begin{align}\label{ChPTK}
{\cal{M}}(K^+\to\pi^+\pi^0)&=\left[\frac{5}{3}G_{27}f_{\pi}(m_K^2-m_\pi^2)-f_{\pi}\delta m^2\left(G_8+\frac{3}{2}G_{27}\right)\right]e^{i\delta_0^2}\nonumber\\
&\sim \frac{5}{3}G_{27}f_{\pi}(m_K^2-m_\pi^2)e^{i\delta_0^2}\equiv {\cal{M}}_K e^{i\delta_0^2},
\end{align}
and\footnote{Regarding $G_8$ and $G_{27}$, in this work we adopt the conventions of Ref.~\cite{D'Ambrosio:1996nm}.}
\begin{align}
F_1^{(DE)}&=-\frac{2ie G_8e^{i\delta_1^1}}{f_{\pi}}\left\{ q\cdot p_2{\cal{N}}_E^{(0)}+\frac{2}{3} q^2{\cal{N}}_E^{(1)}+2q^2L_9\right\},\nonumber\\
F_2^{(DE)}&=\frac{2ie G_8e^{i\delta_1^1}}{f_{\pi}}\left\{q\cdot p_1{\cal{N}}_E^{(0)}-\frac{1}{3} q^2{\cal{N}}_E^{(2)}\right\},\nonumber\\
F_3^{(DE)}&=-\frac{2e G_8e^{i\delta_1^1}}{f_{\pi}}{\cal{N}}_M^{(0)},
\end{align}
where $\delta_0^2$ and $\delta_1^1$ are strong phases associated with final state interactions (FSI) and we have defined the following combinations of counterterms:
\begin{align}\label{weakN}
{\cal{N}}_E^{(0)}&=N_{14}-N_{15}-N_{16}-N_{17},\nonumber\\
{\cal{N}}_E^{(1)}&=N_{14}-N_{15},\nonumber\\
{\cal{N}}_E^{(2)}&=N_{14}+2N_{15}-3(N_{16}-N_{17}),\nonumber\\
{\cal{N}}_M^{(0)}&=\frac{1}{8\pi^2}\bigg[2+3(2a_3-a_2)\bigg].
\end{align}
Bremsstrahlung is a leading ${\cal{O}}(p^2)$ effect while direct emission appears as a subleading ${\cal{O}}(p^4)$ effect. Notice that in the direct emission we have only included the counterterms: since the loop diagrams have no absorptive pieces their contribution can be absorbed in the $N_i$ constants. The results above are analogous to the ones in $K^+\to\pi^+\pi^0\gamma$, except for the off-shell pieces in $F_{1,2}$. We note that ${\cal{N}}_E^{(1)}$ above is also encountered in $K^+\to\pi^+e^+e^-$, while ${\cal{N}}_E^{(2)}$ is a genuine contribution to $K^+\to\pi^+\pi^0e^+e^-$. These off-shell effects, as we will see later, turn out to be sizeable and make, even for rather small values of $q^2$, dramatic (and interesting) differences between $K^+\to\pi^+\pi^0\gamma$ and $K^+\to\pi^+\pi^0\gamma^*$.   
 

\section{Dalitz plot analysis}\label{sec:IV}

Here we will proceed in a similar way as it has already been done for $K^+\to \pi^+\pi^0\gamma$, where an accurate Dalitz plot analysis in the variables $(E_{\gamma }^{\ast }, T_{c}^{\ast })$ was necessary to disentangle the Bremsstrahlung contribution, $\Gamma _{B}$, from the direct emission electric ($E_{DE}$) and magnetic ($M_{DE}$) amplitudes. In that case one can write~\cite{D'Ambrosio:1992bf,Cappiello:2007rs}
\begin{align}
\frac{d^{2}\Gamma (K^+ \to   \pi ^+ \pi^0  \gamma) }{dT_{c}^{\ast }dE_{\gamma }^{\ast }} &=
\frac{d^{2}\Gamma _{B}}{dT_{c}^{\ast }dE_{\gamma }^{\ast }}
\left[1+ 2{\mathrm{Re}}\left(\frac{E_{DE}}{e{\cal M}_K}\right) \rho+\left( \left| \frac{E_{DE}}{e{\cal M}_K}\right| ^{2}+\left| \frac{M_{DE}}{e{\cal M}_K}\right| ^{2}\right) \rho^2\right], \label{eq:TcEg}
\end{align}
with
\begin{equation}\label{elmag}
E_{DE}=-\frac{2eG_8m_K^3}{f_{\pi}}{\cal{N}}_E^{(0)};\qquad M_{DE}=-\frac{2eG_8m_K^3}{f_{\pi}}{\cal{N}}_M^{(0)},
\end{equation}
and
\begin{align}
\rho&=\left[E_{\gamma}^*+T_c^*+\left(m_{\pi}-\frac{m_K}{2}\right)\left(1-\frac{\delta m^2}{2m_Km_{\pi}}\right)\right]\frac{E_\gamma^*}{m_K}=\left( \frac{m_{K}}{2}-E_{0}-\frac{\delta m^2}{2m_K}\right)\frac{E_{\gamma
}^{\ast }}{m_{K}}.
\end{align}
$E_{0}$ above is the $\pi ^0$-energy in the kaon rest frame and ${\cal M}_K$ is defined in Eq.~(\ref{ChPTK}). In compliance with Low's theorem, the Bremsstrahlung contribution can be factored out and one can extract the electric and magnetic parts from the $\rho$ dependence. In this way NA48/2 obtained a determination of $E_{DE}$ from the interference term in Eq.~(\ref{eq:TcEg}) together with $|M_{DE}|$~\cite{:2010uja}. $E_{DE}$ and $M_{DE}$ give, respectively, dynamical information on the non-anomalous and anomalous $p^4$ weak chiral lagrangian combinations ${\cal{N}}_E^{(0)}$ and ${\cal{N}}_M^{(0)}$, which can be compared with existing theoretical predictions~\cite{Ecker:1992de,D'Ambrosio:1997tb,Cappiello:2011re}. In particular, it would be interesting to check independently the interference term, since NA48/2 finds opposite sign compared to the one expected on theoretical grounds. Moreover, $K^+\to\pi^+\pi^0\gamma$ has no reach over the sign of $M_{DE}$, which has not been determined yet. 

Here we want to discuss $K^{\pm}\to\pi^{\pm}\pi^0e^+e^-$, complementing the kinematical variables used in Eq.~(\ref{eq:TcEg}) with the dilepton invariant mass, $q^2$. As already emphasized in the Introduction, the $K^+\to\pi^+\pi^0e^+e^-$ decay rate is largely dominated by the Bremsstrahlung contribution $(BR=4.2\cdot10^{-6})$, followed by the magnetic term (70 times smaller) and the electric interference (120-130 times smaller). The pure electric term is much suppressed (40 to 100 times smaller than the magnetic). While sheer numbers show that the Bremsstrahlung dominates, this dominance is however not homogeneous in the kinematic variables. In Figure~3 we show the $q$ dependence of the different contributions while Table~1 shows the integrated decay rates (the area under the curves) for different kinematic cuts in $q$. The general trend is that the Bremsstrahlung dominance gets reduced as $q^2$ increases. However, even more information can be extracted if one looks at the distribution in the full $(E_{\gamma}^*, T_c^*, q^2,\theta_{\ell}, \phi)$ hyperplane. The angular dependences are purely kinematic and can thus be integrated out. Eventually, the dynamically relevant objects we will focus on are the differential decay rates
\begin{align}\label{dqet}
\frac{d^3\Gamma}{dE_{\gamma}^{\ast}dT_c^{\ast}dq^2}&=\frac{d^3\Gamma_{B}}{dE_{\gamma}^{\ast}dT_c^{\ast}dq^2}+\frac{d^3\Gamma_E}{dE_{\gamma}^{\ast}dT_c^{\ast}dq^2}+\frac{d^3\Gamma_M}{dE_{\gamma}^{\ast}dT_c^{\ast}dq^2}+\frac{d^3\Gamma_{int}}{dE_{\gamma}^{\ast}dT_c^{\ast}dq^2},
\end{align}
where $\Gamma_{int}$ collects the different interference contributions, BE, BM and EM. The decay rates above can be easily computed from
\begin{align}
d^3\Gamma_i&=\frac{1}{2m_K}\sum_{{\mathrm{spins}}} |{\cal M}_{LD}|_i^2 d^3\Phi,
\end{align}
where $d^3\Phi$ is the angular-integrated invariant phase space, whose explicit expression is given in Eq.~(\ref{ps3}). The matrix elements for the different contributions can be readily inferred from Eq.~(\ref{squared}). We note that $\Gamma_{int}$ above actually consists only of the BE term. This is so because the remaining (electric-magnetic) interferences are P-violating, {\it{i.e.}} odd in $\phi$. Thus, both BM and EM terms cancel upon angular integration ({\it{cf.}} Section~\ref{sec:VI} for a full discussion of the electric-magnetic interferences.)
\begin{figure}[t]
\begin{center}
\includegraphics[width=11.5cm]{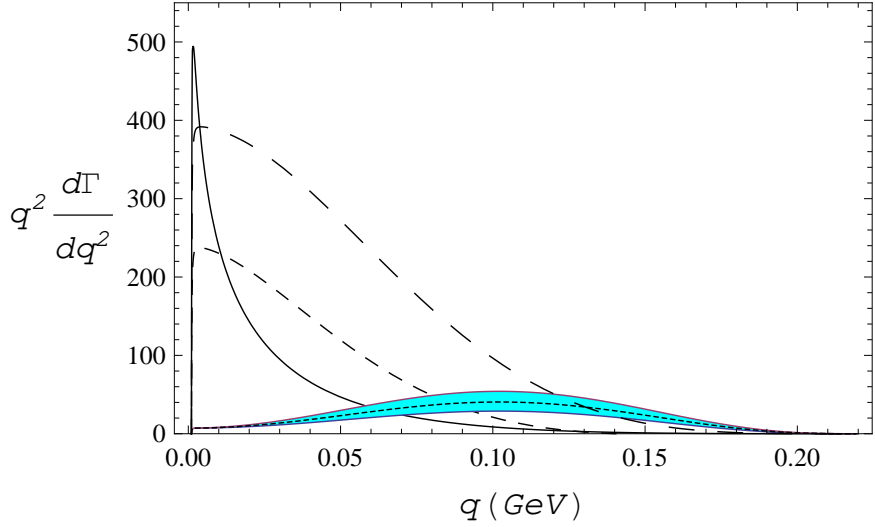}
\end{center}
\caption{\small{\it{$q$ dependence of the different contributions. The solid line represents the Bremsstrahlung. The dashed lines (from bigger dash to smaller dash) are $100\times$M, $100\times$BE and $300\times$E, respectively. Error bars are omitted. The band around the electric contribution corresponds to varying ${\cal{N}}_E^{(1,2)}$ in the range $\pm{\cal{N}}_E^{(0)}$. The corresponding band on the electric interference term is completely negligible.}}}
\end{figure}
\begin{table}[t]
\begin{center}
\begin{tabular}{|cccccc|}
\hline
$q_c$ (MeV)&\,\,\,\,\, B $[10^{-8}]$ &\,\,\,\,\, B/M&\,\,\,\,\, B/E&\,\,\,\,\,B/BE&\,\,\,\,\,B/BM\\
\hline
$2m_l$ &\,\,\,\,\, 418.27 &\,\,\,\,\, 71&\,\,\,\,\, 4405&\,\,\,\,\, 128&\,\,\,\,\, 208 \\
$2$ &\,\,\,\,\, 307.96 &\,\,\,\,\, 61&\,\,\,\,\, 3416&\,\,\,\,\, 111&\,\,\,\,\, 165\\
$4$ &\,\,\,\,\, 194.74 &\,\,\,\,\, 48&\,\,\,\,\, 2320&\,\,\,\,\, 90&\,\,\,\,\, 129\\
$8$ &\,\,\,\,\, 109.60 &\,\,\,\,\, 36&\,\,\,\,\, 1414&\,\,\,\,\, 71&\,\,\,\,\, 100\\
$15$ &\,\,\,\,\, 56.12 &\,\,\,\,\, 26&\,\,\,\,\, 789&\,\,\,\,\, 56&\,\,\,\,\, 78\\
$35$ &\,\,\,\,\, 15.50 &\,\,\,\,\, 16&\,\,\,\,\, 263&\,\,\,\,\, 41&\,\,\,\,\, 54\\
$55$ &\,\,\,\,\, 5.62 &\,\,\,\,\, 12&\,\,\,\,\, 118&\,\,\,\,\, 38&\,\,\,\,\, 44\\
$85$ &\,\,\,\,\, 1.37 &\,\,\,\,\, 9&\,\,\,\,\, 46&\,\,\,\,\, 49&\,\,\,\,\, 37\\
$100$ &\,\,\,\,\, 0.67 &\,\,\,\,\, 8&\,\,\,\,\, 30&\,\,\,\,\, 71&\,\,\,\,\, 36\\
$120$ &\,\,\,\,\, 0.24 &\,\,\,\,\, 8&\,\,\,\,\, 18&\,\,\,\,\, 458&\,\,\,\,\, 35\\
$140$ &\,\,\,\,\, 0.04 &\,\,\,\,\, 9&\,\,\,\,\, 10&\,\,\,\,\, -45&\,\,\,\,\, 37\\
$180$ &\,\,\,\,\, 0.003 &\,\,\,\,\, 12&\,\,\,\,\, 5&\,\,\,\,\, -19&\,\,\,\,\, 44\\
\hline
\end{tabular}
\end{center}
\caption{\small{\it{Branching ratios for the Bremsstrahlung and the relative weight of the rest of the contributions for different cuts in $q$, starting at $q_{min}$ (first row) and ending at $180$ MeV. In the last column we have also included the parity-odd magnetic-electric inteference term, to be discussed in Section~\ref{sec:VI}. }}}\label{tab1}
\end{table}

For instance, with the definitions for the form factors $F_i^{(B)}$ and $F_3^{(DE)}$ and the kinematic weights of Eq.~(\ref{Tkinematic}), and upon using Eq.~(\ref{conversion}), one can easily show that the explicit expressions for the Bremsstrahlung and magnetic terms read (in the isospin limit):   
\begin{align}\label{results}
\frac{d^3\Gamma_{B}}{dE_{\gamma}^*dT_c^*dq^2}&=\frac{\alpha^2|{\cal{M}}_K|^2(2m_l^2+q^2)}{48\pi^3 m_K^3q^4}\sqrt{1-\frac{4m_l^2}{q^2}}\frac{q^6+\lambda_1q^4+\lambda_2q^2-4m_K^2\sigma}{{(q^2-2E_{\gamma}^*m_K)^2(m_K-2(m_{\pi}+T_c^*+E_{\gamma}^*))^2}},\nonumber\\
\frac{d^3\Gamma_{M}}{dE_{\gamma}^*dT_c^*dq^2}&=-\frac{\alpha^2 G_8^2m_K({\cal{N}}_M^{(0)})^2(2m_l^2+q^2)}{48\pi^3f_{\pi}^2q^4}\sqrt{1-\frac{4m_l^2}{q^2}}\bigg\{q^4+(2\lambda_3-\lambda_4)q^2+\sigma\bigg\},
\end{align}
where
\begin{align}\label{deflambda}
\lambda_1&=2m_K(m_K-4E_{\gamma}^*-2m_{\pi}-2T_c^*)-4m_{\pi}^2,\nonumber\\
\lambda_2&=m_K^2(16E_{\gamma}^{*2}-8E_{\gamma}^*m_K+(m_K-2(m_{\pi}+T_c^*))(m_K+6(m_{\pi}+T_c^*)))+16m_{\pi}^2m_KE_{\gamma}^*,\nonumber\\
\lambda_3&=(m_K-2m_{\pi})(m_K-2E_{\gamma}^*)-2E_{\gamma}^*m_{\pi},\nonumber\\
\lambda_4&=4T_c^*[m_K-E_{\gamma}^*-2m_{\pi}-T_c^*],\nonumber\\
\sigma&=\lambda_3^2-m_K(m_K-2E_{\gamma}^*)\lambda_4,\nonumber\\
\delta&\equiv \delta_1^1-\delta_0^2,
\end{align}
Similarly, expressions for the remaining electric and electric interference terms can be found.

In order to extract useful information from the previous expressions it is instructive to study the shape of each contribution in the $(E_{\gamma}^*, T_c^*)$ Dalitz plot while scanning over $q^2$. Thus, for any fixed $q^2$ we will have the corresponding Dalitz plot. This is a very convenient strategy to uncover how the different contributions populate the Dalitz plot and how this picture evolves dynamically. Throughout our analysis we will use the following set of parameter values:
\begin{align}
m_K&=493.677\,\, {\mathrm{MeV}};& m_{\pi}&=137.293\,\, {\mathrm{MeV}};& m_{\ell}&=0.511\,\, {\mathrm{MeV}};\nonumber\\ \alpha &=(137.036)^{-1};& f_{\pi}&=93\,\, {\mathrm{MeV}};& G_8&=9.0\cdot 10^{-6}\,\,{\mathrm{GeV}}^{-2};\nonumber\\
L_9&=6.9\cdot 10^{-3};& {\cal{N}}_E^{(0)}&=-2.2164\cdot 10^{-3};& {\cal{N}}_M^{(0)}&=2.85\cdot 10^{-2};\nonumber\\
&& \delta&\equiv\delta_1^1-\delta_0^2\sim\frac{\pi}{18}.&&
\end{align}
The values for ${\cal{N}}_E^{(0)}$ and ${\cal{N}}_M^{(0)}$ have been taken from Ref.~\cite{:2010uja}. As mentioned in the Introduction, the experimental value for ${\cal{N}}_E^{(0)}$ turns out to be negative, in contrast to the theoretical estimates given in Refs.~\cite{Ecker:1992de,D'Ambrosio:1997tb}, which yield a number comparable in magnitude but with positive sign. We want to point out that the holographic approach taken in Ref.~\cite{Cappiello:2011re} actually yields ${\cal{N}}_E^{(0)}=0$. This confirms that strong cancellations are a characteristic of ${\cal{N}}_E^{(0)}$ and, therefore, that it is difficult to ascertain its value from hadronic models. The remaining weak counterterms ${\cal{N}}_E^{(1,2)}$, in the absence of an experimental value, could in principle be estimated using hadronic models. However, a reliable estimate would require to evaluate the pion and kaon loops that we implicitly absorbed in the counterterms ${\cal{N}}_E^{(1,2)}$, something that we will not do here. Instead, we will vary them in the range {\mbox{${\cal{N}}_E^{(1,2)}\in [-{\cal{N}}_E^{(0)},{\cal{N}}_E^{(0)}]$.}} 
 
In Figure~4 we show snapshots of the different contributions at $q^2=(50$ MeV$)^2$. One can observe that the different contributions populate quite distinct regions of the Dalitz plot. At very low values of $q^2$ the BE piece is almost filling the same spot as the Bremsstrahlung, while the direct emission pieces stay in the center. This is precisely what one finds in $K^+\to\pi^+\pi^0\gamma$ and it is thus a cross-check that our results in the on-shell photon limit $q^2\to0$ are correct. However, even for modest values of $q^2$ the different contributions drift towards different edges of the Dalitz plot, whith the exception of the magnetic term, which remains centered. The picture shown in Figure~4 remains qualitatively valid up until values of $q^2$ close to the endpoint. As Table~1 illustrates, the rather distant but still looming Bremsstrahlung contribution diminishes its relative strength as $q^2$ increases. However, the overall signal to background ratio also drops sharply with $q^2$, so for experimental detection large $q^2$ values are statistically disfavored. A fiducial regime can be found however at $q^2=(50$ MeV$)^2$ or even smaller. This is possible because of the sharp evolution with $q^2$ of the different contributions. 

We want to finish this Section by giving a quantitative estimate of the expected experimental signals for all the contributions over the Bremsstrahlung, when one optimizes the information encoded in the $(E_{\gamma}^*,T_c^*)$ Dalitz plot and concentrates on the regions where each contribution is most favored. Tables~2-4 show the expected ratios for BE, electric and magnetic contributions, respectively. The numbers quoted are the result of the integrated decay rates over a square centered at the peak of each contribution. Since the shape of the Dalitz plot changes with running $q^2$ ({\it{cf.}} the $q^2$ dependence of $E_{\gamma}^*$ and $T_c^*$ in Eq.~(\ref{limits})), the squares are conventionally normalized such that their area is always $1/8$ times the full Dalitz plot area.\footnote{We chose a square for simplicity. Any other shape (with the same area) would lead to similar results.} The comparison is then done by integrating the same region for the remaining contributions.

Table~2 shows the integrated Dalitz plot around the maximum of the electric interference term for different cuts in $q^2$. In order to see the improvement that can be achieved by placing cuts on $E_{\gamma}^*$ and $T_c^*$ over a naive Dalitz plot integration, it is useful to compare with the results of Table~1. Notice that around the optimal cut $q^2=(50$ MeV$)^2$, the electric interference term is 30 times smaller than the Bremsstrahlung, which is comparable to what one finds in Table~1. However, electric and magnetic backgrounds have dropped significantly (from $25\%$ to $2\%$ for the electric and a factor 4 for the magnetic). Comparison of Tables~1 and 3 shows that the electric signal gets magnified by one order of magnitude with respect to the Bremsstrahlung. The electric intereference background drops by roughly a factor 5, while, because of its proximity, no significant reduction of the magnetic background is achieved. Regarding the magnetic contribution, since the electric and interference backgrounds were already small, the important point is that it gets enhanced by a factor 2 with respect to the Bremsstrahlung.

In summary, a much cleaner extraction of the interference and the magnetic terms is possible: in both cases a significant enhancement of the signal is achieved (over the Bremsstrahlung) with small backgrounds from the remaining contributions. With this enhanced determination of the magnetic term one can even attempt an extraction of the electric piece: the signal is enhanced while the interference background gets reduced.    

As we mentioned above, we have neglected isospin breaking effects in our analysis. However, notice that, since the Bremsstrahlung is the dominant contribution, isospin breaking effects there can constitute as sizeable an effect as the rest of the contributions. We have checked that isospin effects are sizeable only close to the Bremsstrahlung peak and therefore do not invalidate our previous conclusions. For completeness we list in the Appendix the explicit expression for the Bremsstrahlung contribution away from the isospin limit.
 
\begin{table}[t]
\begin{center}
\begin{tabular}{|ccccc|}
\hline
$q_c$ (MeV) &\,\,\,\,\, B/BE &\,\,\,\,\, BE&\,\,\,\,\, E/BE&\,\,\,\,\,M/BE\\
\hline
10 &\,\,\,\,\, $701.90$&\,\,\,\,\, $1.6\cdot 10^{-22}$ &\,\,\,\,\, $5.4\cdot 10^{-4}$&\,\,\,\,\, $0.07$\\
30 &\,\,\,\,\, $165.34$&\,\,\,\,\, $2.6\cdot 10^{-23}$ &\,\,\,\,\, $2.3\cdot 10^{-3}$&\,\,\,\,\, $0.27$\\
50 &\,\,\,\,\, $29.05$&\,\,\,\,\, $1.0\cdot 10^{-23}$ &\,\,\,\,\, $0.02$&\,\,\,\,\, $1.23$\\
70 &\,\,\,\,\, $40.75$&\,\,\,\,\, $2.5\cdot 10^{-24}$ &\,\,\,\,\, $0.01$&\,\,\,\,\, $0.72$\\
100 &\,\,\,\,\, $11.96$&\,\,\,\,\, $6.9\cdot 10^{-25}$ &\,\,\,\,\, $0.04$&\,\,\,\,\, $0.60$\\
120 &\,\,\,\,\, $7.07$&\,\,\,\,\, $3.2\cdot 10^{-25}$ &\,\,\,\,\, $0.07$&\,\,\,\,\, $0.41$\\
\hline
\end{tabular}
{\caption{\small{\it{Branching ratios in a square centered at the maximum of the electric interference contribution (second column) for different cuts in $q$ starting at $q_{min}$ (first row) and ending at $180$ MeV. In the remaining columns we show the relative weight of the rest of the contributions.}}}}\label{tab3}
\end{center}
\end{table}

\begin{table}[t]
\begin{center}
\begin{tabular}{|ccccc|}
\hline
$q_c$ (MeV) &\,\,\,\,\, B/E &\,\,\,\,\, BE/E&\,\,\,\,\, E&\,\,\,\,\,M/E\\
\hline
10 &\,\,\,\,\, $733.48$&\,\,\,\,\, $49.26$ &\,\,\,\,\, $5.9\cdot 10^{-24}$&\,\,\,\,\, $154.68$\\
30 &\,\,\,\,\, $34.22$&\,\,\,\,\, $4.32$ &\,\,\,\,\, $7.0\cdot 10^{-25}$&\,\,\,\,\, $30.66$\\
50 &\,\,\,\,\, $18.44$&\,\,\,\,\, $0.71$ &\,\,\,\,\, $4.7\cdot 10^{-25}$&\,\,\,\,\, $13.13$\\
70 &\,\,\,\,\, $7.02$&\,\,\,\,\, $-0.48$ &\,\,\,\,\, $3.1\cdot 10^{-25}$&\,\,\,\,\, $5.27$\\
100 &\,\,\,\,\, $3.57$&\,\,\,\,\, $-1.15$ &\,\,\,\,\, $1.7\cdot 10^{-25}$&\,\,\,\,\, $2.14$\\
120 &\,\,\,\,\, $3.39$&\,\,\,\,\, $-1.74$ &\,\,\,\,\, $1.1\cdot 10^{-25}$&\,\,\,\,\, $1.37$\\
\hline
\end{tabular}
{\caption{\small{\it{Branching ratios in a square centered at the maximum of the Electric contribution (third column) for different cuts in $q$ starting at $q_{min}$ (first row) and ending at $180$ MeV. In the remaining columns we show the relative weight of the rest of the contributions.}}}}\label{tab4}
\end{center}
\end{table}

\begin{table}[t]
\begin{center}
\begin{tabular}{|ccccc|}
\hline
$q_c$ (MeV) &\,\,\,\,\, B/M &\,\,\,\,\, BE/M&\,\,\,\,\, E/M&\,\,\,\,\,M\\
\hline
10 &\,\,\,\,\, $5.48$&\,\,\,\,\, $0.34$ &\,\,\,\,\, $6.4\cdot 10^{-3}$&\,\,\,\,\, $9.1\cdot 10^{-22}$\\
30 &\,\,\,\,\, $5.21$&\,\,\,\,\, $0.33$ &\,\,\,\,\, $9.6\cdot 10^{-3}$&\,\,\,\,\, $8.8\cdot 10^{-23}$\\
50 &\,\,\,\,\, $4.81$&\,\,\,\,\, $0.30$ &\,\,\,\,\, $0.02$&\,\,\,\,\, $2.5\cdot 10^{-23}$\\
70 &\,\,\,\,\, $4.46$&\,\,\,\,\, $0.26$ &\,\,\,\,\, $0.03$&\,\,\,\,\, $9.2\cdot 10^{-24}$\\
100 &\,\,\,\,\, $4.33$&\,\,\,\,\, $0.10$ &\,\,\,\,\, $0.07$&\,\,\,\,\, $2.2\cdot 10^{-24}$\\
120 &\,\,\,\,\, $5.30$&\,\,\,\,\, $0.06$ &\,\,\,\,\, $0.10$&\,\,\,\,\, $7.8\cdot 10^{-25}$\\
\hline
\end{tabular}
{\caption{\small{\it{Branching ratios in a square centered at the maximum of the Magnetic contribution (fourth column) for different cuts in $q$ starting at $q_{min}$ (first row) and ending at $180$ MeV. In the remaining columns we show the relative weight of the rest of the contributions.}}}}\label{tab5}
\end{center}
\end{table} 
  

\section{SM short-distance contributions to $K^\pm\to\pi^\pm\pi^0e^+e^-$}\label{sec:V}

The effective Hamiltonian for $\Delta S=1$ transitions reads~\cite{Dib:1988js,Buchalla:1995vs}:  
\beq
{\cal H}_{eff}^{|\Delta S| = 1} = -\frac{G_F}{\sqrt{2}} \,V_{ts}^* V_{td} \Big[ \, y_{7\gamma}(\mu) )
Q_{7\gamma}(\mu) ~+~ y_{7A}(m_W) Q_{7A}(m_W) \, \Big]\, +\, \mbox{h.c.},
\eeq
where $V_{ij}$ denote CKM matrix elements and
\begin{equation}
Q_{7\gamma} \, = (m_s{\bar s}_R\sigma^{\mu\nu}d_L+m_d{\bar s}_L\sigma^{\mu\nu}d_R)
F_{\mu\nu}, \qquad \quad
Q_{7A} \, =  \, \overline{s}_L \gamma^{\mu} d_L \, \overline{\ell}
\gamma_{\mu} \gamma_5 \ell,
\label{eq:q7v7a}
\end{equation}
with ${\displaystyle L,R=\frac{1}{2}(1\mp\gamma_5)}$ and ${F_{\mu\nu}=-i(q_{\mu}\epsilon_{\nu}-q_{\nu}\epsilon_{\mu})}$. Below we will consider each operator in more detail. 

The first operator induces a CP-violating $sd\gamma^*$ vertex of the form~\cite{Dib:1988js,Buchalla:1995vs,Riazuddin:1993pn}
\begin{align}
{\cal H}_{\it eff}^{\gamma}&={\sqrt 2}G_F\frac{ie}{(4\pi)^2}V_{ts}^*V_{td}
\ D' (x_t)Q_{7\gamma}
+ {\rm h.c.},
\label{eq:heff}
\end{align}
where
\begin{align}
D' (x_t)=\left(\frac{x_t}{2(x_t-1)^3}\right)
\left(\frac{8x_t^2+5x_t-7}{6}-\frac{x_t(3x_t-2)\ln{x_t}}{(x_t-1)}\right),\qquad {x_t=m_t^2/M_W^2}.
\label{eq:Dprime}
\end{align}
It is a well-known fact~\cite{Riazuddin:1993pn} that the chiral realization of the hadronic matrix element in Eq.~(\ref{eq:heff}) requires a dimensionally-suppressed operator at ${\cal O}(p^6)$. However, under renormalization this operator mixes with $Q_-$ and its Wilson coefficient gets enhanced~\cite{Dib:1990gr}.\footnote{We note that even stronger enhancements can happen in BSM scenarios, {\it{e.g.}} supersymmetry~\cite{Colangelo:1999kr,Tandean:2000qk}.} This operator will be reconsidered later on in Section~\ref{sec:VI} when we discuss CP violation.

$Q_{7A}$ is induced by W-box and Z-penguin diagrams. In the presence of such contributions the matrix element for $K^+\to\pi^+\pi^0e^+e^-$ takes the generic expression
\begin{align}
{\cal{M}}_{SD}=[{\bar{u}}(k_-)\gamma^{\mu}\gamma_5 v(k_+)]{\cal{H}}_{\mu}(p_1,p_2,q),
\end{align}
where the hadronic tensor ${\cal{H}}_{\mu}$ is 
\begin{align}
{\cal H}^\mu(p_1,p_2,q)=F_1^{\rm SD} p_1^\mu+F_2^{\rm SD} p_2^\mu+F_3^{\rm SD} \varepsilon^{\mu\nu\alpha\beta}p_{1\nu}p_{2\alpha}q_{\beta}.
\end{align}
The interference between this axial leptonic current and the long-distance $\gamma^*$-contribution discussed in previous Sections will give interesting P-violating effects that we will analyze in Sect.~\ref{sec:VI}.

In order to evaluate the contributions to the form factors $F_i^{\rm SD}$ above, one has to consider the bosonized version of the axial-vector effective operator
\begin{align}
{\cal H}_{\it SD}=\xi \frac{s_1 G_F \alpha}{\sqrt{2}}\bar s \gamma_\mu (1 - \gamma_5)
d\   \bar e \gamma^\mu \gamma_5 e +{\rm{h.c.}},  \label{eq:Hshdis}
\end{align} 
where $s_1$ is a short-hand notation for the sine of the Cabibbo angle. The quantity $\xi$ receives significant contributions from both top and charm quark loops and is given by~\cite{Savage:1990km,Lu:1992eqa} 
\begin{align}
\xi = -\tilde{\xi}_c + \left(\frac{V_{ts}^* V_{td}}{V_{us}^* V_{ud}}\right)\tilde{\xi}_t,
\end{align}
where
\begin{align}
\tilde{\xi}_q=\tilde{\xi}_q^{(Z)}+ \tilde{\xi}_q^{(W)},
\end{align}
have been calculated including perturbative QCD corrections at the next-to-leading order. $\tilde \xi_c$ is of order $10^{-4}$ while $\tilde \xi_t$ is of order unity~\cite{Buchalla:1992zm,Buchalla:1993bv}.

The chiral realization of the hadronic current in Eq.~(\ref{eq:Hshdis}) is, at the lowest order~\cite{deRafael:1995zv}:   
\begin{align}
\bar{s}\gamma_\mu(1-\gamma_5)d\rightarrow -f_{\pi}^2(L_\mu)_{23}+\frac{i}{8\pi^2}\varepsilon_{\mu\nu\alpha\beta}(L^\nu L^\alpha L^{\beta})_{23},
\end{align}
where $L_\mu=iU^{\dagger}\partial_\mu U$. Thus, at very low energies one has
\begin{align}
{\cal H}_{SD} =i\xi\frac{G_F \alpha s_1}{2\sqrt{2}}\left[f_{\pi}^2\,  {\mathrm{Tr}} (U^{\dagger}\partial^\mu U \lambda_6)-\frac{i}{8\pi^2}\varepsilon^{\mu\nu\lambda\rho}\,{\mathrm{Tr}}(U^{\dagger}\partial_{\nu}U\partial_{\lambda}U^{\dagger}\partial_{\rho}U\lambda_6)\right] \bar e \gamma_\mu \gamma_5 e+{\rm{h.c.}}
\end{align}
Expanding out the $U$ field above one easily obtains the following expression for the decay amplitude: 
\begin{align}
{\cal{M}}_{SD}=-i\xi \frac{G_F \alpha s_1}{2 \sqrt{2} f_{\pi}}\left[p_1^\mu - p_2^\mu-\frac{i}{2\pi^2f_{\pi}^2}\varepsilon^{\mu\nu\lambda\rho}p_{1\nu}p_{2\lambda}q_{\rho}
\right]\bar u (k_-) \gamma_\mu \gamma_5 v(k_+),
\end{align}
which immediately implies that
\begin{align}
F_1^{\rm SD}=-F_2^{\rm SD}=2if_{\pi}^2\pi^2F_3^{\rm SD},\label{rel}\\
F_1^{\rm SD}=-i\xi \frac{G_F\alpha s_1}{2\sqrt{2}f_{\pi}}.\label{relSM}
\end{align}
Notice that the three form factors are related by trivial numerical coefficients. This statement holds also when strong phases are taken into account: FSI in the previous form factors are described only by $\delta_1^1$. Thus, in $K^+\to\pi^+\pi^0e^+e^-$, $\delta_0^2$ is exclusive of the Bremsstrahlung contribution while $\delta_1^1$ is the only strong phase that affects short distances. 
\newpage
\begin{figure}[t]
\begin{center}
\includegraphics[width=6.0cm]{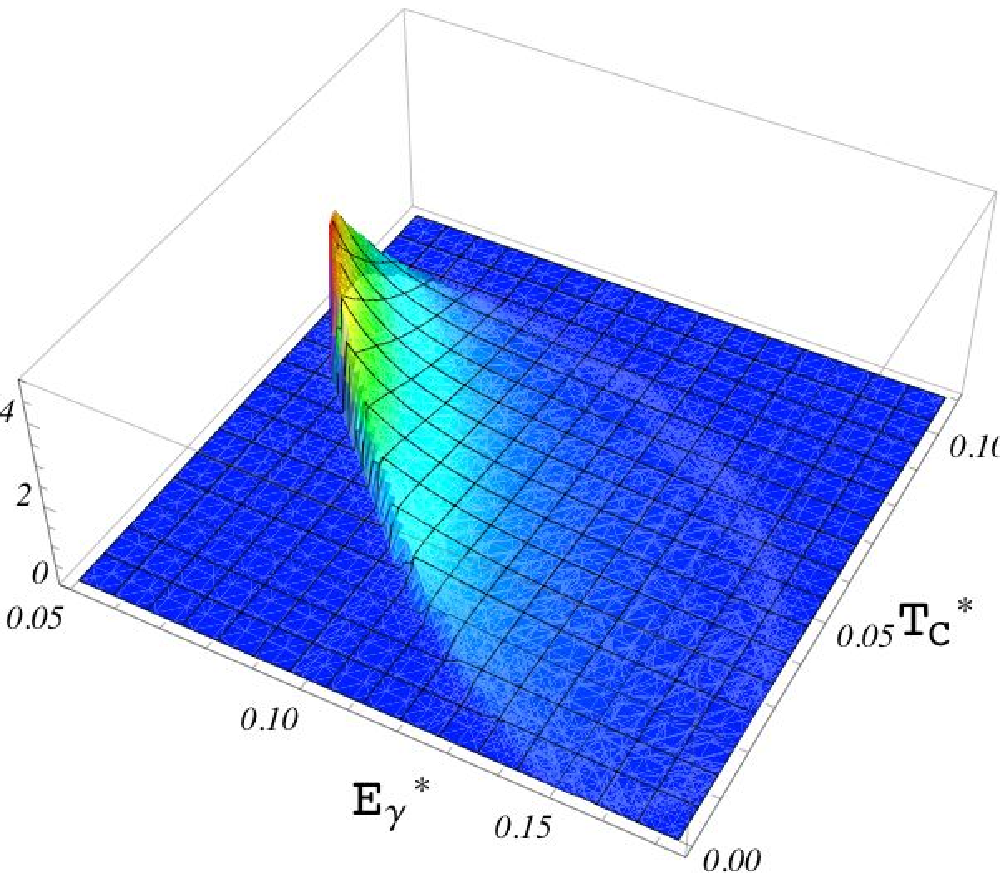}\hskip 1.0cm \includegraphics[width=7.0cm]{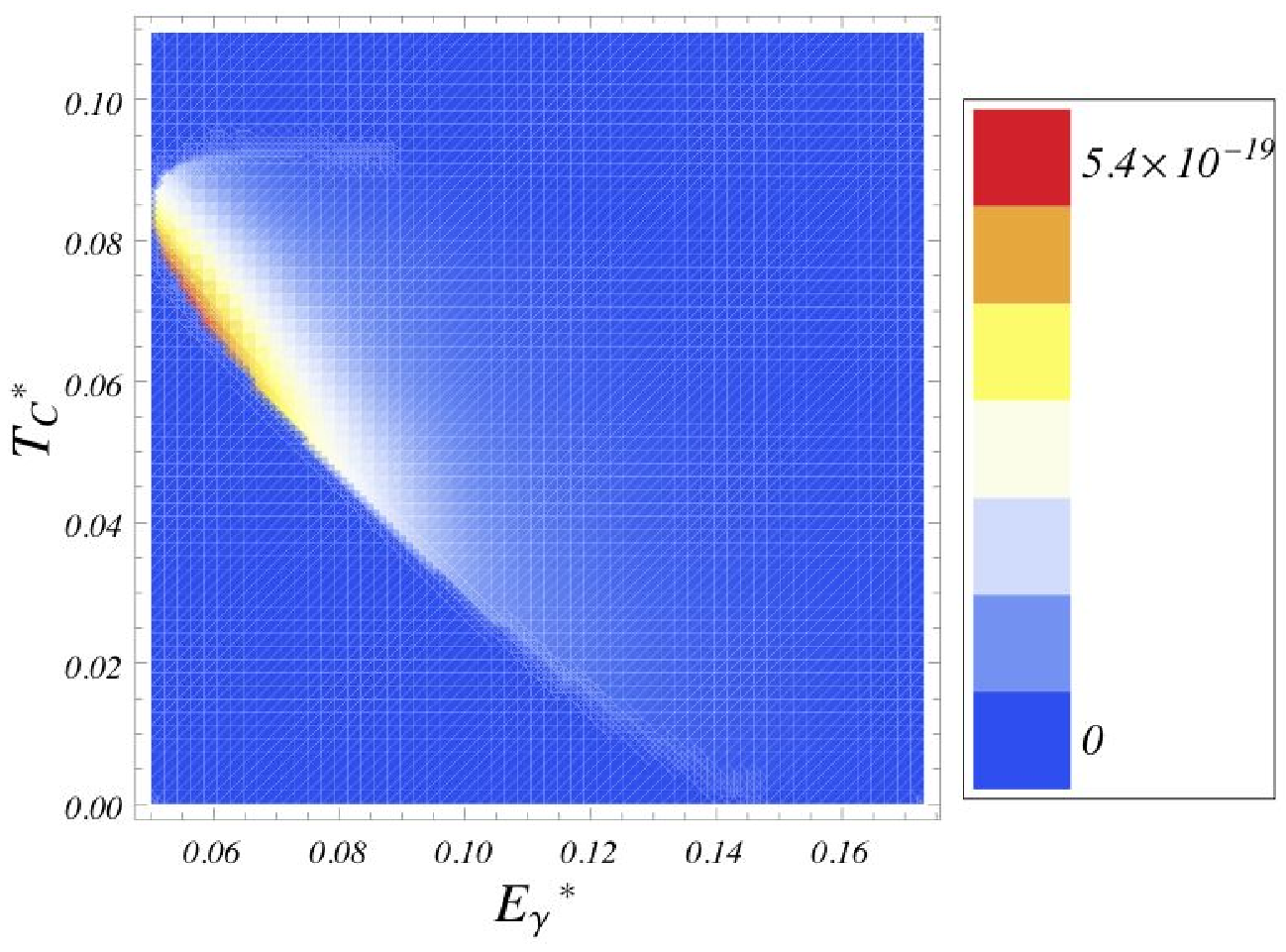}
\vskip 0.2cm
\includegraphics[width=6.0cm]{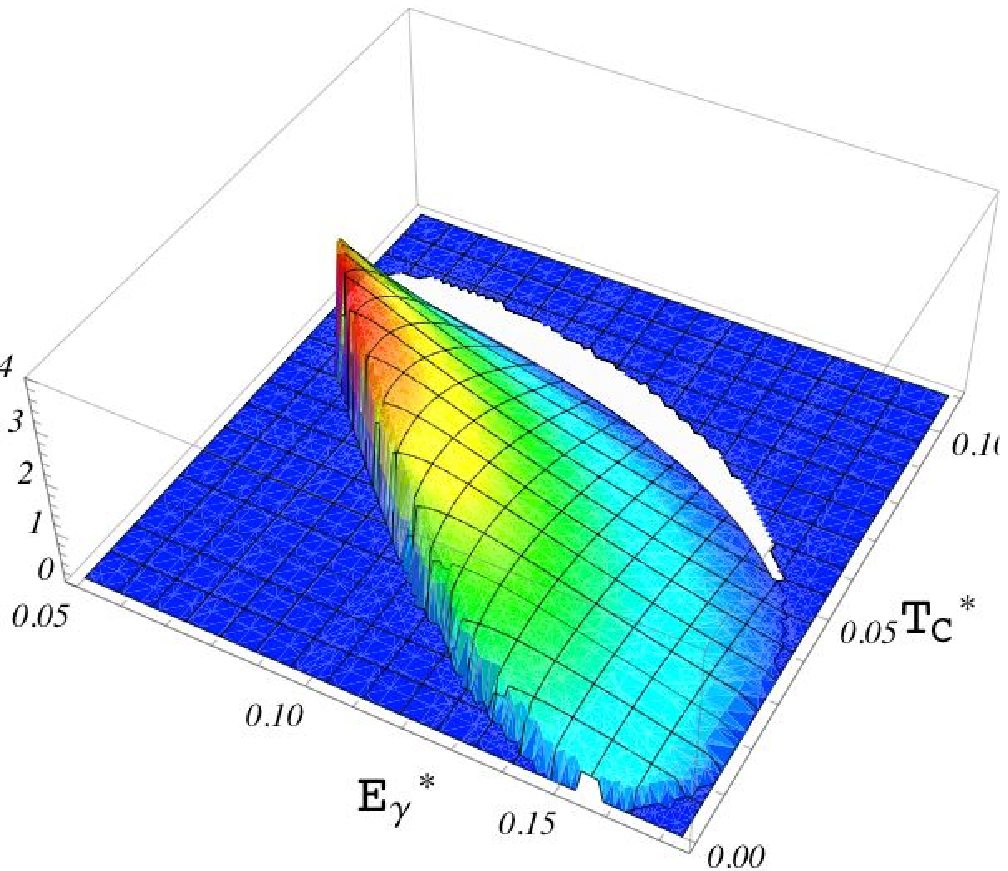}\hskip 1.0cm \includegraphics[width=7.0cm]{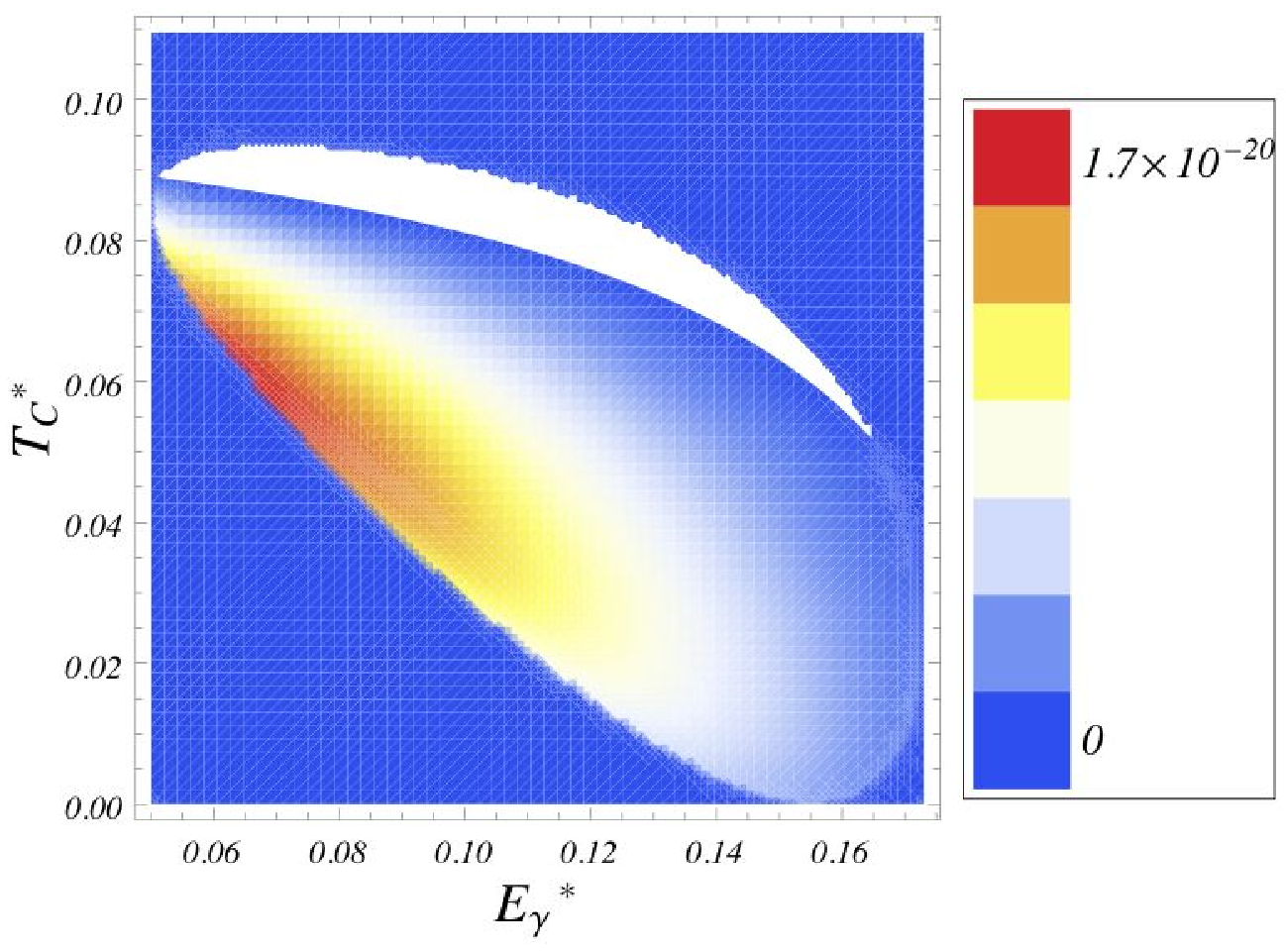}
\vskip 0.2cm
\includegraphics[width=6.0cm]{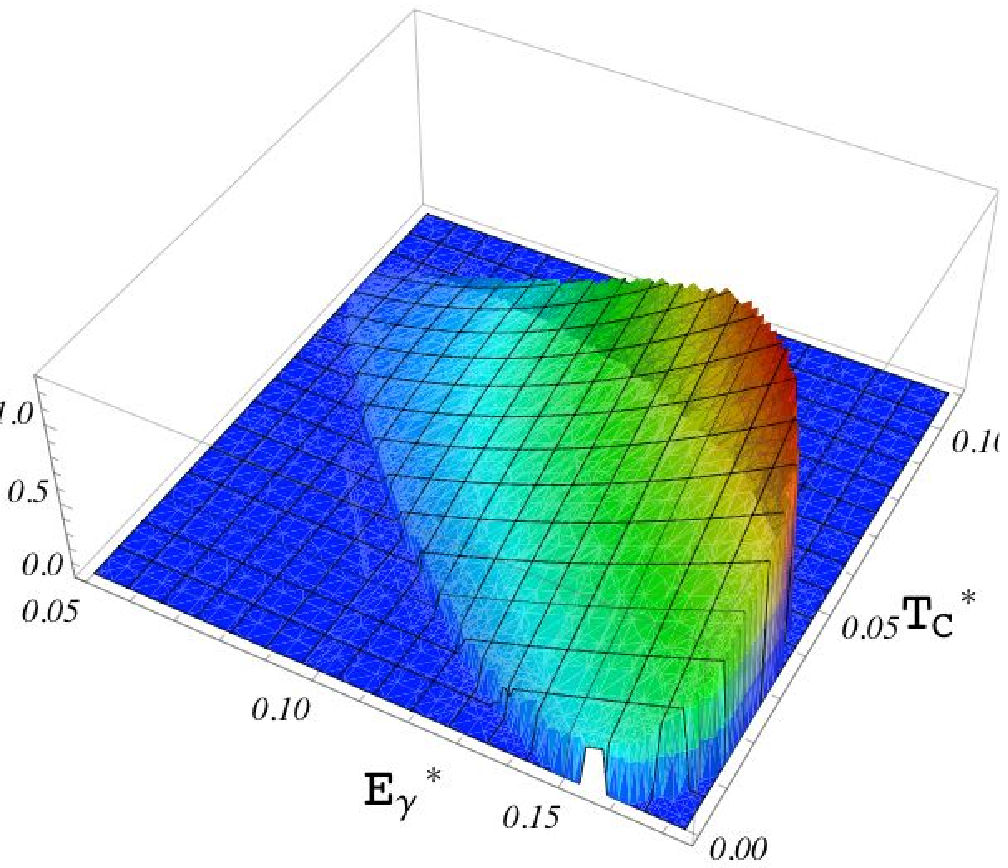}\hskip 1.0cm \includegraphics[width=7.0cm]{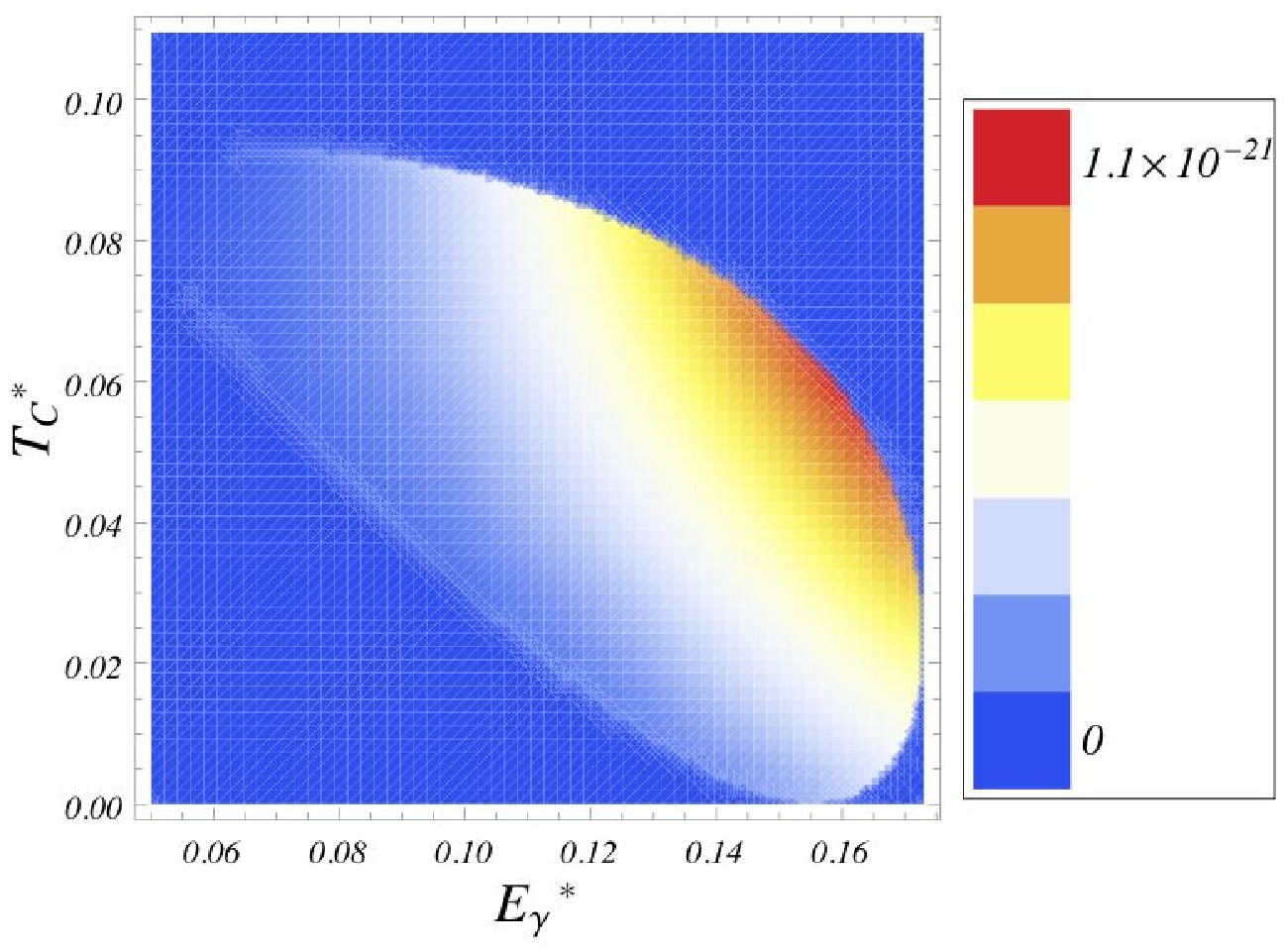}
\vskip 0.2cm
\includegraphics[width=6.0cm]{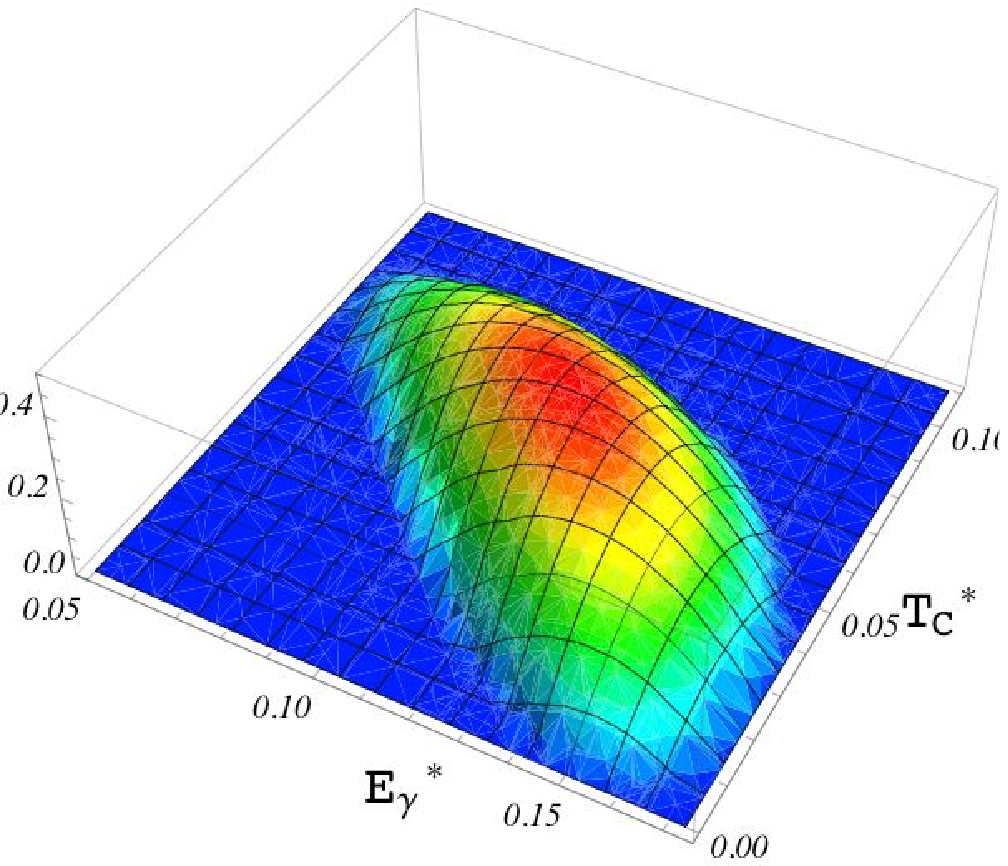}\hskip 1.0cm \includegraphics[width=7.0cm]{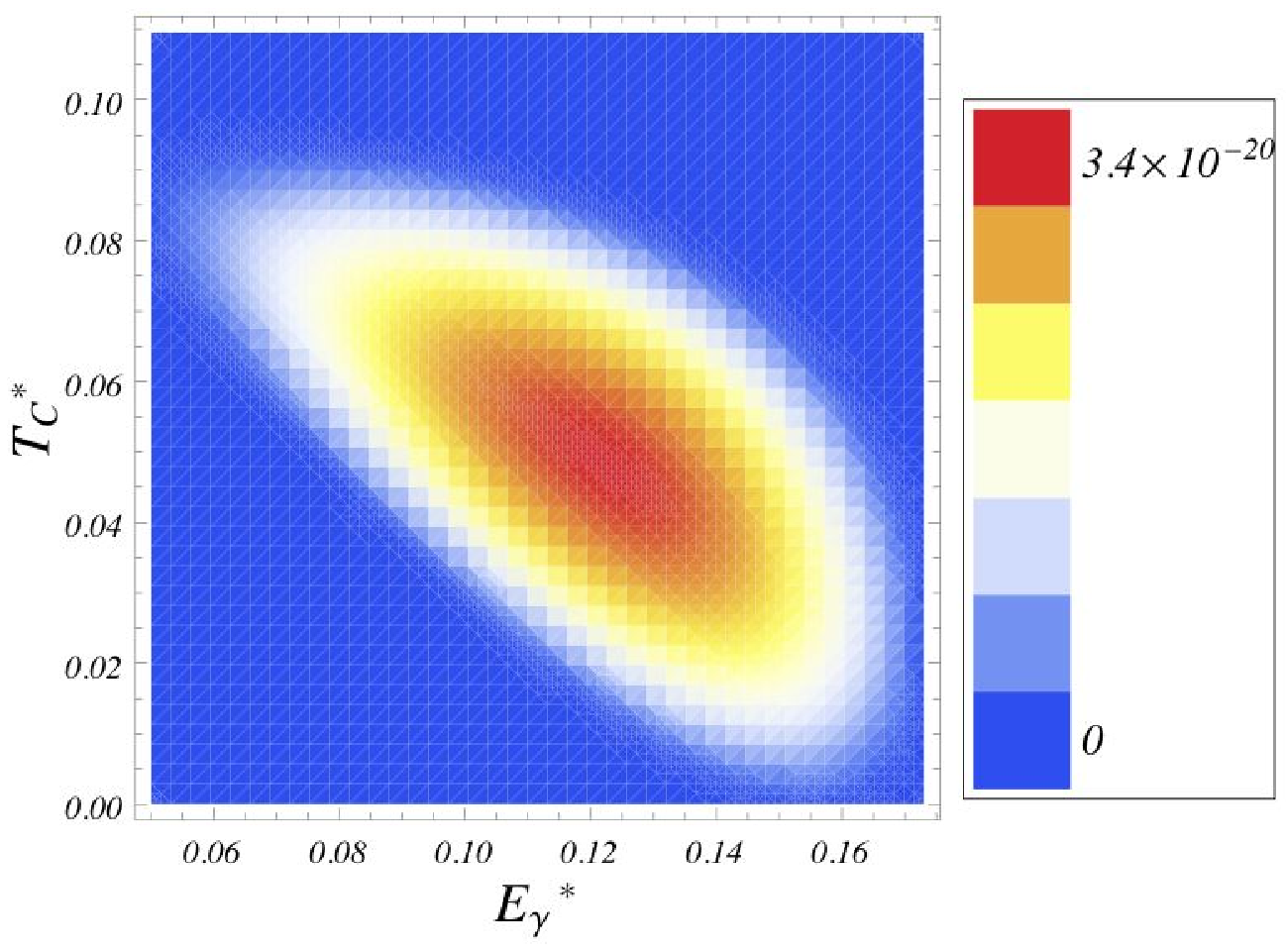}
\end{center}
\caption{\small{\it{On the left panels: Dalitz plot in the $(E_{\gamma}^{\ast},T_c^{\ast})$ plane at $q^2=(50$ MeV$)^2$ for the different contributions: B, BE, E and M (from top to bottom). On the right panels we show their two-dimensional density projections.}}}
\end{figure} 
\newpage

 
\section{P- and  CP-violation in $K^+\to\pi^+\pi^0e^+e^-$ }\label{sec:VI}

In the previous Sections we have concentrated on the (dominant) long-distance photon-mediated contribution to $K^+\to\pi^+\pi^0e^+e^-$ as well as short-distance SM effects. In this Section we want to ascertain which asymmetries can be used to extract the different sources of short-distance physics. In order to carry out this short-distance analysis it will prove convenient to express the differential rate in the general form given for $K_{l4}$ decays~\cite{Cabibbo:1965zz,Pais:1968zz}:
\begin{align}\label{angular}
\frac{d^5\Gamma}{dE_{\gamma}^*dT_c^*dq^2d\cos\theta_{\ell} d\phi}&={\cal{A}}_1+{\cal{A}}_2\sin^2\theta_{\ell}+{\cal{A}}_3\sin^2\theta_{\ell}\cos^2\phi+{\cal{A}}_4\sin2\theta_{\ell}\cos\phi\nonumber\\
&+{\cal{A}}_5 \sin\theta_{\ell}\cos\phi+{\cal{A}}_6 \cos\theta_{\ell}+{\cal{A}}_7 \sin\theta_{\ell}\sin\phi\nonumber\\
&+{\cal{A}}_8\sin 2\theta_{\ell}\sin\phi+{\cal{A}}_9\sin^2\theta_{\ell}\sin 2\phi,
\end{align}
where $\theta_{\ell}$ and $\phi$ are two of the Cabibbo-Maksymowicz angular variables for $K_{l4}$ decays~\cite{Cabibbo:1965zz,Bijnens:1994me} and ${\cal{A}}_i$ are dynamical functions that can be parameterized in terms of 3 dynamical variables, {\it{e.g.}}, ${\cal{A}}_i(E_{\gamma}^*,T_c^*,q)$. This factorized form makes it transparent which angular asymmetries are required to isolate the different ${\cal{A}}_i$. 

One can easily show that the Bremsstrahlung, direct emission and electric interference terms contribute to ${\cal{A}}_{1-4}$. In contrast, ${\cal{A}}_{8,9}$ receive contributions from the electric-magnetic interference terms (BM and EM) and therefore capture long-distance induced P-violating terms. ${\cal{A}}_{5,6,7}$ are also P-violating terms but generated through the interference of $Q_{7A}$ with long distances.

In this Section we will extract information about ${\cal{A}}_{5-9}$ through a set of angular asymmetries. We will thus discuss P violation both at long and short distances, but also the extraction of CP-violating signals from $K^+\to\pi^+\pi^0e^+e^-$. 


\subsection{P violation at long distances}\label{asymmLD}

While doing the Dalitz plot analysis in Section~\ref{sec:III} we mentioned that the magnetic interference term cancels after a full angular integration. This is due to the intrinsic parity-odd nature of the magnetic term. Actually, integrating Eq.~(\ref{angular}) with respect to $\theta_{\ell}$, the differential cross section looks like:
\begin{equation}
\frac{d\Gamma}{d\phi}={\cal{I}}_1\cos^2\phi+{\cal{I}}_2\sin^2\phi+{\cal{I}}_3\sin\phi\cos\phi+{\cal{I}}_4\sin\phi+{\cal{I}}_5\cos\phi,
\end{equation}
where ${\cal{I}}_{1,2}$ are P-conserving and include the B, E, M and BE contributions while ${\cal{I}}_{3,4,5}$ are P-violating. In this Section we will be interested in ${\cal{I}}_3$, which consists entirely of the BM contribution and thus probes P violation at long distances\footnote{The EM contribution, also inside ${\cal{I}}_3$, is negligible.}. In contrast, ${\cal{I}}_{4,5}$ are related to ${\cal{A}}_{5,6,7}$ and therefore test P violation at short distances ({\it{cf.}} the next Section).  

One way to isolate ${\cal{I}}_3$ is to define an appropriate asymmetry in $\phi$. Since this term is asymmetric with respect to the shift $\phi\to\pi-\phi$, if one defines the following piece-wise angular integration 
\begin{equation}
\int_0^{2\pi}d\phi^*\equiv \left[\int_0^{\pi/2}-\int_{\pi/2}^{\pi}+\int_{\pi}^{3\pi/2}-\int_{3\pi/2}^{2\pi}\right]d\phi,
\end{equation}
then the P-violating observable 
\begin{align}\label{Plong}
A_P^{(L)}&=\frac{\displaystyle\int_0^{2\pi}\frac{d\Gamma}{d\phi}d\phi^*}{\displaystyle\int_0^{2\pi}\frac{d\Gamma}{d\phi}d\phi},
\end{align}
naturally selects the magnetic interference piece. Indeed, the contributions for ${\cal{I}}_1$, ${\cal{I}}_2$, ${\cal{I}}_4$ and ${\cal{I}}_5$ identically cancel above and we are left with
\begin{equation}
\frac{d^3\Gamma_{BM}}{dE_{\gamma}^*dT_c^*dq^2}=-\frac{\alpha^2G_8{\cal{M}}_K{\cal{N}}_M^{(0)}}{24\pi^4f_{\pi}m_K q^2}\left(1-\frac{4m_l^2}{q^2}\right)^{3/2}\frac{q^4+(2\lambda_3-\lambda_4)q^2+\sigma}{\sqrt{E_{\gamma}^{*2}-q^2}(m_K-2(E_{\gamma}^*+m_{\pi}+T_c^*))}\sin\delta,
\end{equation}
where ${\cal{M}}_K$, ${\cal{N}}_E^{(0)}$ and ${\cal{N}}_M^{(0)}$ are defined in Eqs.~(\ref{ChPTK}) and (\ref{weakN}), while $\lambda_3$, $\lambda_4$, $\sigma$ and $\delta$ are given in Eqs.~(\ref{deflambda}). Incidentally, we want to note that the structure entering the numerator coincides with the one found in the magnetic term (see Eq.~(\ref{results})). 

The observable defined in Eq.~(\ref{Plong}) was already used to extract the magnetic interference term in $K_L\to\pi^+\pi^-e^+e^-$~\cite{Lai:2003ad,Abouzaid:2005te}. However, in that case the ${\cal{O}}(p^4)$ contribution is strongly suppressed and ${\cal{O}}(p^6)$ terms dominate. In $K^+\to\pi^+\pi^0e^+e^-$, in contrast, the ${\cal{O}}(p^4)$ is already sizeable and counterterms in ${\cal{N}}_M^{(0)}$ are expected to be small. A measurement of $A_P^{(L)}$ can confirm the sign of the magnetic interference and thus the convergence of the chiral expansion. Notice that the previous contribution can also be used to extract the strong phase $\delta$. 

In Figure~5 we show the Dalitz plot at the representative $q=50$ MeV cut. We note that the shape of the Dalitz plot for this contribution is largely independent of $q^2$ (this is to be expected because neither the magnetic nor the Bremsstrahlung pieces vary much). Table~5 shows the comparison of the expected signal against the (Bremsstrahlung-dominated) background. As discussed in previous sections ({\it{cf.}} also Table~1), $q=50$ MeV is probably the optimal cut: lower values of $q^2$ are largely dominated by the Bremsstrahlung while higher values are afflicted with poor statistics. Therefore, naively (without better cuts), the precision required to extract the BM interference is of the same order needed to isolate the magnetic term in $K^+\to\pi^+\pi^0e^+e^-$. 

\begin{figure}[t]
\begin{center}
\includegraphics[width=6.0cm]{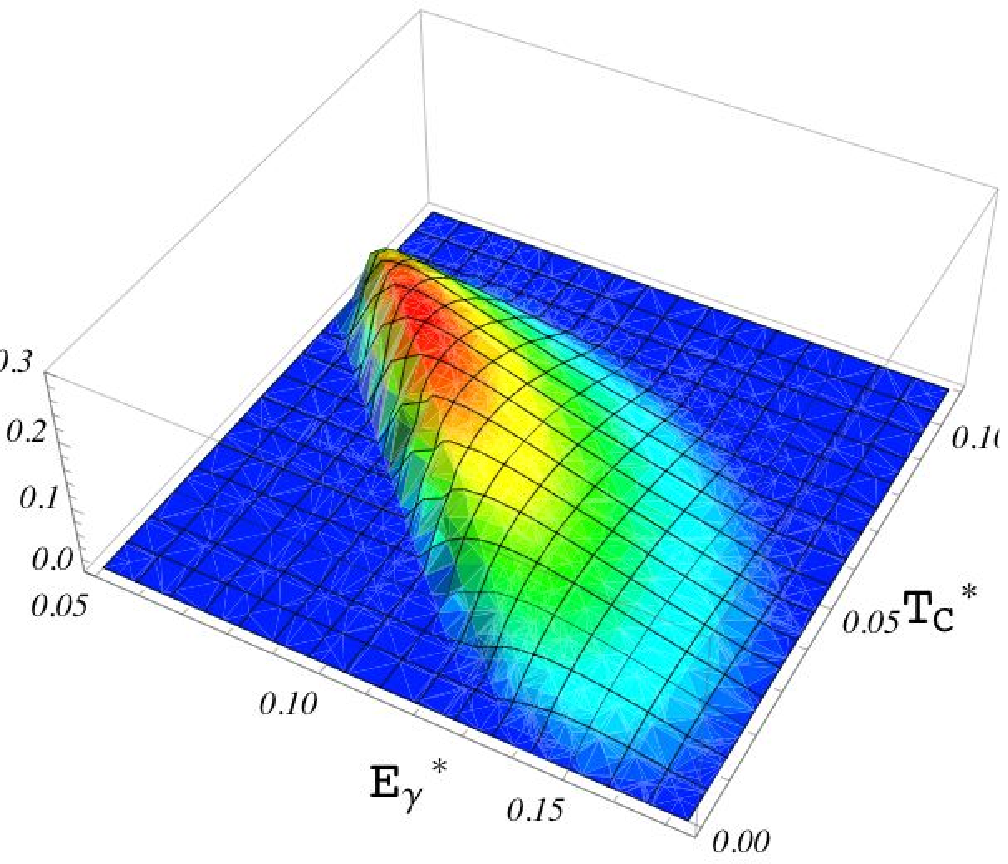}\hskip 1.0cm \includegraphics[width=7.0cm]{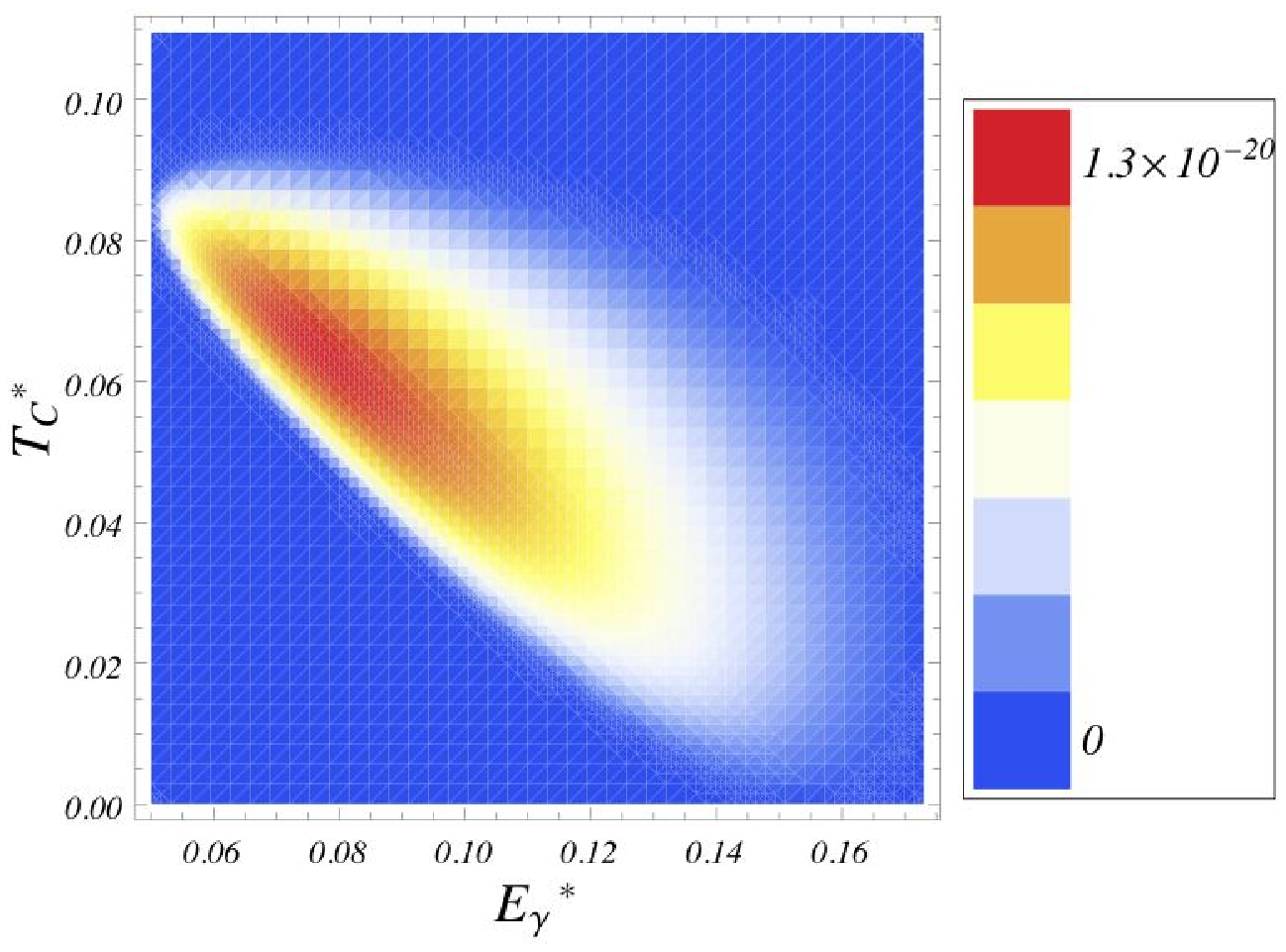}
\end{center}
\caption{\small{\it{On the left panel: Dalitz plot in the $(E_{\gamma}^{\ast},T_c^{\ast})$ plane at $q^2=(50$ MeV$)^2$ for the P-violating BM contribution, obtained after a piece-wise angular integration. On the right panel we show the corresponding two-dimensional density projection.}}}
\end{figure}

\begin{table}[t]
\begin{center}
\begin{tabular}{|cccccc|}
\hline
$q_c$ (MeV) &\,\,\,\,\, B/BM &\,\,\,\,\, BE/BM&\,\,\,\,\, E/BM&\,\,\,\,\,M/BM&\,\,\,\,\,BM\\
\hline
10 &\,\,\,\,\, $593.03$&\,\,\,\,\, $1.57$ &\,\,\,\,\, $2.0\cdot 10^{-3}$&\,\,\,\,\, $0.31$&\,\,\,\,\,  $2.6\cdot 10^{-22}$\\
30 &\,\,\,\,\, $150.75$&\,\,\,\,\, $1.56$ &\,\,\,\,\, $6.6\cdot 10^{-3}$&\,\,\,\,\, $0.88$&\,\,\,\,\, $2.9\cdot 10^{-23}$\\
50 &\,\,\,\,\, $85.01$&\,\,\,\,\, $1.45$ &\,\,\,\,\, $1.5\cdot 10^{-2}$&\,\,\,\,\, $1.46$&\,\,\,\,\, $7.6\cdot 10^{-24}$\\
70 &\,\,\,\,\, $58.77$&\,\,\,\,\, $1.20$ &\,\,\,\,\, $3.4\cdot 10^{-2}$&\,\,\,\,\, $2.02$&\,\,\,\,\, $2.5\cdot 10^{-24}$\\
100 &\,\,\,\,\, $40.29$&\,\,\,\,\, $0.55$ &\,\,\,\,\, $0.12$&\,\,\,\,\, $2.78$&\,\,\,\,\, $5.4\cdot 10^{-25}$\\
120 &\,\,\,\,\, $33.90$&\,\,\,\,\, $-0.13$ &\,\,\,\,\, $0.26$&\,\,\,\,\, $3.17$&\,\,\,\,\,$1.9\cdot 10^{-25}$\\
\hline
\end{tabular}
{\caption{\small{\it{Branching ratios in a square centered at the maximum of the magnetic interference contribution (fifth column) for different cuts in $q$ starting at $q_{min}$ (first row) and ending at $180$ MeV. In the remaining columns we show the relative weight of the rest of the contributions.}}}}\label{tab6}
\end{center}
\end{table}

\subsection{P violation at short distances}

In this section we would like to isolate P-violating effects that come entirely from short distances. This can be done by considering P violation not in the hadronic tensor but in the dilepton pair. This way one ensures that there is no contamination from the long-distance QED background~\cite{Heiliger:1993qt}. We have already noted in Section~\ref{sec:V} that axial leptonic vertices, which are generated by $W$-box and $Z$-penguin diagrams, induce P-violating effects through interference with vector leptonic currents. In this Section we will use this fact to study P violation both within and beyond the SM. 

Similar to Section~\ref{sec:V} we will consider the generic expression:
\begin{align}
{\cal{M}}_{SD}=[{\bar{u}}(k_-)\gamma^{\mu}\gamma_5 v(k_+)]{\cal{H}}_{\mu}(p_1,p_2,q),
\end{align}
where ${\cal{H}}_{\mu}$ takes the form,
\begin{align}\label{Hshort}
{\cal{H}}^{\mu}={\cal{F}}_1 p^\mu_1+{\cal{F}}_2 p_2^\mu+{\cal{F}}_3 \varepsilon^{\mu\nu\alpha\beta}p_{1\nu}p_{2\alpha}q_\beta.
\end{align}
The form factors ${\cal{F}}_i$ contain short distance information, namely the W and Z one-loop contributions evaluated in Section~\ref{sec:V} as well as potential contributions from physics beyond the SM. With the normalization we have chosen, ${\cal{F}}_i=F_i^{SD}+{\cal{F}}_i^{BSM}$. 

The dominant P-violating piece will arise from the interference between the long-distance (photon-mediated) matrix element ${\cal{M}}_{LD}$ and the short-distance ${\cal{M}}_{SD}$ with an axial dilepton current. 
The resulting amplitude reads
\begin{align}
{\cal{M}}_{LD}{\cal{M}}_{SD}^*+{\cal{M}}_{SD}{\cal{M}}_{LD}^*&=\frac{8e}{q^2}{\mathrm{Im}}\left[H^{\mu}{\cal{H}}^{\nu\,*}\right]\epsilon_{\mu\nu\lambda\rho}k_+^{\lambda}k_-^{\rho}=\frac{4e}{q^2}\sum_{i<j}G_{ij}{\mathrm{Im}}[{\cal{F}}_i{{F}}_j^*-{\cal{F}}_j{{F}}_i^*], \label{eq:EMinterf}
\end{align}
where ${{F}}_i$ are the form factors defined in Section~\ref{sec:III} and $G_{ij}$ are given by
\begin{align}
G_{12}&=\epsilon_{\mu\nu\lambda\rho}p_1^{\mu}p_2^{\nu}Q^{\lambda}q^{\rho},\nonumber\\
G_{13}&=-\left[p_1\cdot Q(q^2p_1\cdot p_2-p_1\cdot qp_2\cdot q)+p_2\cdot Q((p_1\cdot q)^2-m_{\pi^+}^2q^2)
\right],\nonumber\\
G_{23}&=\left[p_2\cdot Q(q^2p_2\cdot p_1-p_2\cdot qp_1\cdot q)+p_1\cdot Q((p_2\cdot q)^2-m_{\pi^0}^2q^2)
\right].
\end{align}
As we remarked in the Introduction, the form factors are functions only of the variables $E_{\gamma}^{\ast}$, $T_c^{\ast}$ and $q^2$, while the angular dependence is entirely contained in the kinematical weights $G_{ij}$. Therefore, even without knowing ${\cal{F}}_i$ explicitly, we can readily isolate the angular dependence of the vector-axial interference term as
\begin{align}\label{angular1}
\frac{d^2\Gamma}{d\cos\theta_{\ell} d\phi}&={\cal{A}}_5 \sin\theta_{\ell}\cos\phi+{\cal{A}}_6 \cos\theta_{\ell}+{\cal{A}}_7 \sin\theta_{\ell}\sin\phi,
\end{align}
where $A_{5,6}$ receive contributions from the $G_{13}$ and $G_{23}$ terms in Eq.~(\ref{eq:EMinterf}), while $A_7$ comes entirely from the term proportional to $G_{12}$. Notice from the $G_{ij}$ above that the vector-axial interference is linear in $Q$. An appropriate asymmetry to be sensitive to such contributions is therefore the dilepton momenta asymmetry $k_{\pm}\to-k_{\mp}$. In terms of angles this corresponds to $(\theta_{\ell},\phi)\to(\pi-\theta_{\ell},\pi+\phi)$, and thus a possible observable is
\begin{align}\label{asymSD}
A_P^{(S)}=\frac{\displaystyle\int_0^1 d\cos{\theta_{\ell}}\int_0^{\pi/2}d\phi\frac{d^2\Gamma}{d\phi d\cos{\theta_{\ell}}}-\int_{-1}^{0}d\cos{\theta_{\ell}}\int_{\pi}^{3\pi/2}d\phi\frac{d^2\Gamma}{d\phi d\cos{\theta_{\ell}}}}{\displaystyle\int_0^1 d\cos{\theta_{\ell}}\int_0^{\pi/2}d\phi\frac{d^2\Gamma}{d\phi d\cos{\theta_{\ell}}}+\int_{-1}^{0}d\cos{\theta_{\ell}}\int_{\pi}^{3\pi/2}d\phi\frac{d^2\Gamma}{d\phi d\cos{\theta_{\ell}}}}.
\end{align}  
$A_P^{(S)}$ selects, by construction, short-distance information: BSM or weak interaction couplings of the leptons. For definiteness, and in order to be conservative, we will assume Minimal Flavor Violation (MFV)~\cite{D'Ambrosio:2002ex} for BSM interactions. This means that no new operators other than the SM ones will appear. Consequently, the form factors ${\cal{F}}_i$ are constant and satisfy the relations of Eq.~(\ref{rel}). Notice also that MFV implies the existence of only one weak phase and one strong phase. Taking all this into account, the asymmetry of Eq.~(\ref{asymSD}) takes the form   
\begin{align}\label{Ap}
A_{P}^{(S)}&\sim \frac{e}{\Gamma_Bq^2}\bigg\langle \sum_{i<j}G_{ij}{\mathrm{Im}}[{\cal{F}}_i{{F}}_j^*-{\cal{F}}_j{{F}}_i^*]\bigg\rangle_{(\theta_\ell,\phi)-asym}\nonumber\\
&\sim\frac{e}{\Gamma_Bq^2}\bigg\langle |{\cal{F}}_1|\bigg[G_{12}\sin{\delta}\,|{{F}}^{(B)}|+(G_{13}-G_{23})|M_{DE}|\bigg]\nonumber\\
&\qquad\qquad\qquad\quad+|{\cal{F}}_3|\bigg[G_{13}|{{F}}_1^{(B)}|-G_{23}|{{F}}_2^{(B)}|\bigg]\bigg\rangle_{(\theta_\ell,\phi)-asym},
\end{align}
where ${{F}}^{(B)}={{F}}_1^{(B)}+{{F}}_2^{(B)}$, $M_{DE}$ is given in Eq.~(\ref{elmag}), $\delta$ is the FSI and the brackets stand for phase space integration. Use has been made of Eqs.~(\ref{relSM}) to express the short distance form factors in terms of only ${\cal{F}}_1$ and ${\cal{F}}_3$. This way the magnetic or electric character of the form factors is transparent.

Some comments are in order:
\begin{itemize}
\item Since $K^+$ is not a charge eigenstate, $A_{P}^{(S)}$ contains both CP conserving and CP violating contributions. The latter (proportional to $G_{12}$) are suppressed and therefore not included above (see, however, next Section). 
\item The two terms proportional to ${\cal{F}}_1$ are competitive: ${\cal{F}}_1$ and $M_{DE}$ have a strong phase difference of $90$ degrees and therefore there is no strong phase suppression in the second term. This compensates for the suppression of $|M_{DE}|$ over $|{\hat{F}}^{(B)}|$. However, it is the term proportional to ${\cal{F}}_3$ which eventually dominates $A_{P}^{(S)}$. 
\end{itemize}
In Figure~6 we show the Dalitz plot for each contribution, always assuming MFV, {\it{i.e.}}, constant ${\cal{F}}_i$. Only the shape should be considered meaningful: without specific model predictions for the form factors ${\cal{F}}_i$, the plots have been normalized to arbitrary units. However, they point out the most favored regions to search for each source of P violation. As a matter of fact, the second term in Eq.~(\ref{Ap}) is positive definite, while the first and third terms are not. The plots can be qualitatively understood by noticing that the first and third terms are akin to Bremsstrahlung-electric interference (modulated by the kinematical factors $G_{ij}$), while the second one is a magnetic-electric interference (again modulated by $G_{ij}$). 
\begin{figure}[t]
\begin{center}
\includegraphics[width=6.0cm]{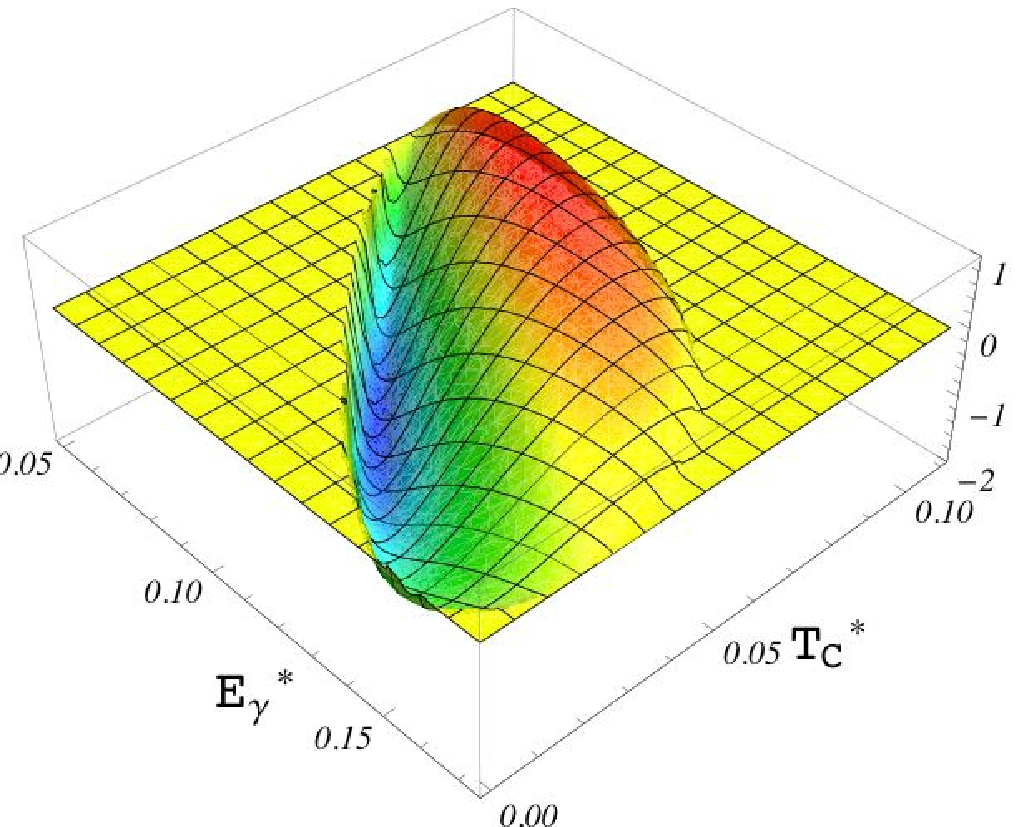}\hskip 1.0cm \includegraphics[width=6.0cm]{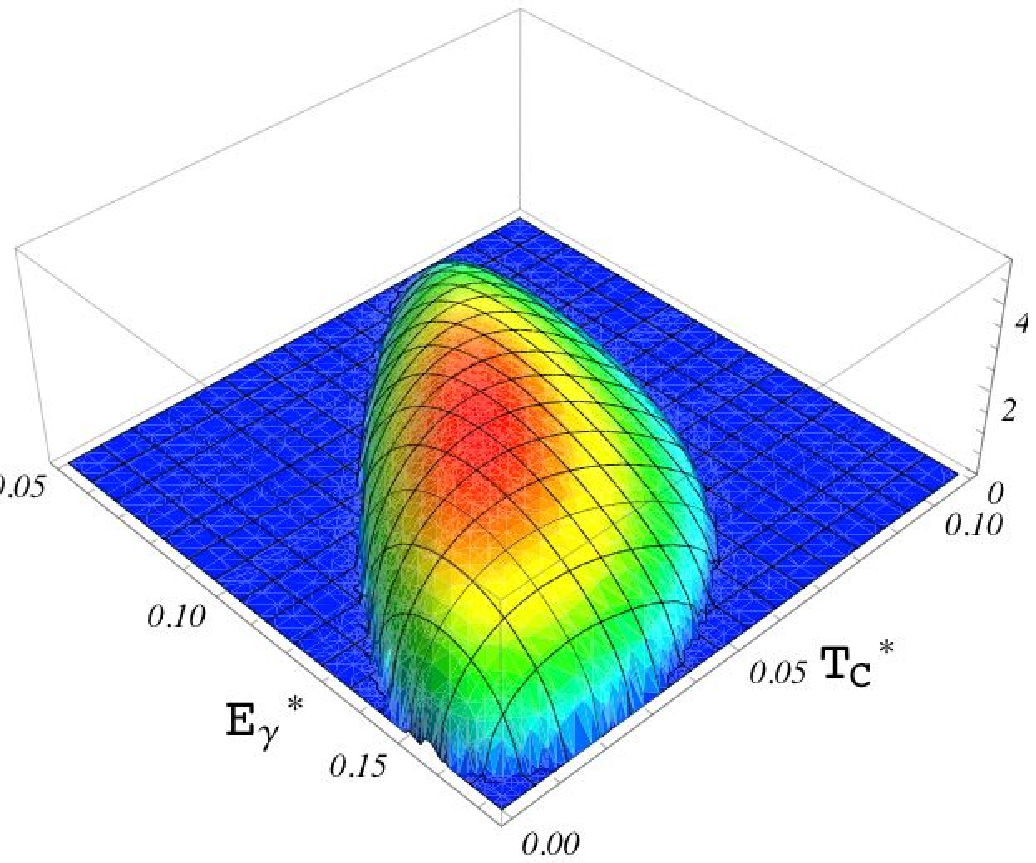}
\newline
\includegraphics[width=6.0cm]{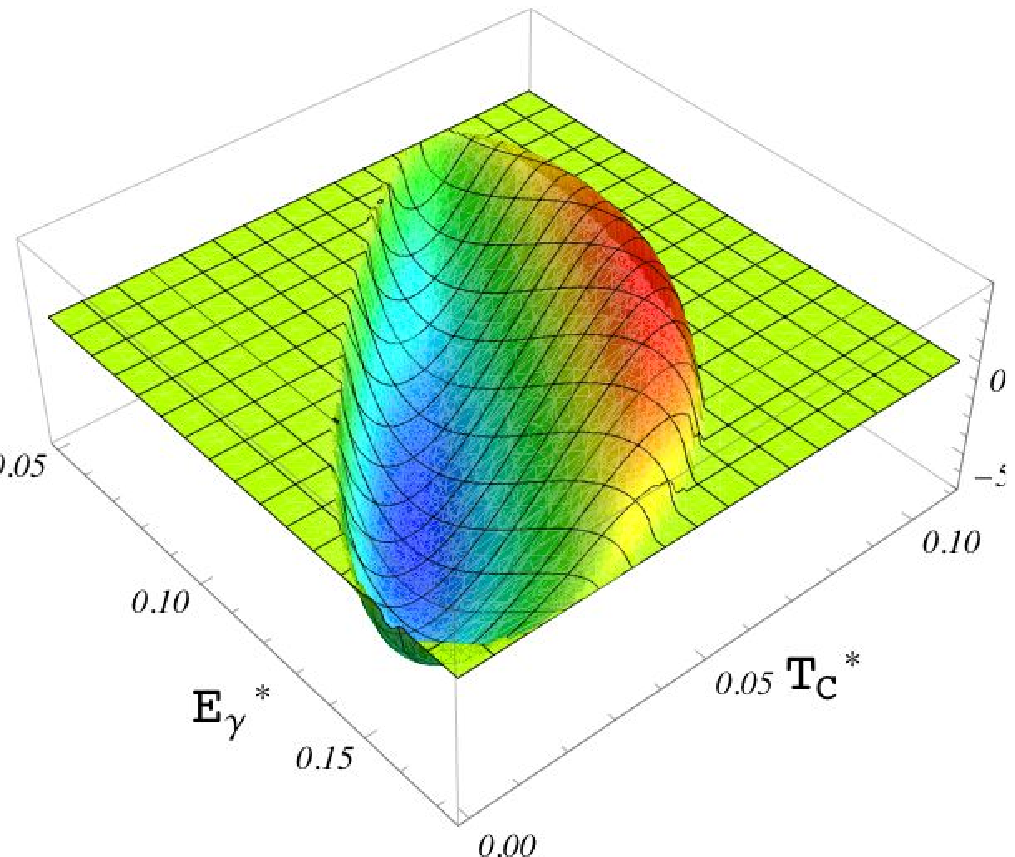}
\caption{\small{\it{Dalitz plot for the contributions to $A_P^{(S)}$ (in arbitrary units) at $q^2=(50$ MeV$)^2$. The different terms are geometrically placed as they appear in Eq.~(\ref{Ap}).}} }
\end{center}
\end{figure}


\subsection{CP violation}

In the previous Section we studied direct contributions to short-distance physics through genuine short-distance operators. One of the conclusions we reached is that new physics can be detected without resorting to charge asymmetries. This is particularly welcomed since no $K^-$ detection is needed to explore short distances. However, we will show that when the charge asymmetry can be built much more short-distance information can be extracted.
 
In order to have a CP-violating charge asymmetry in $K^+\to\pi^+\pi^0e^+e^-$ one has to choose a P-conserving observable. In contrast to the neutral mode $K_L\to\pi^+\pi^- e^+e^-$, here no long distance ($\epsilon$-like) contributions are present and thus charge asymmetry is directly sensitive to short distance effects (electroweak and/or BSM phases). CP-violating phases might show up either associated to new physics operators ({\it{cf}}. last Section) or induced through $K^+\to\pi^+\pi^0\gamma^*$. Therefore, there are two possibilites: i) terms proportional to $e^2$ (P-conservation in {\emph{both}} the leptonic and hadronic pieces) or ii) terms proportional to $e c_A^{\ast}$ (P violation in {\it{both}} the leptonic and hadronic pieces). In this Section we will consider both of them.

The simplest observable one can build is the fully integrated charge asymmetry, which yields the result
\begin{align} 
A_{CP}&=\Frac{\Gamma (K^+ \to \pi^+ \pi^0 e^+ e^-) - \Gamma (K^- \to \pi^- \pi^0 e^+ e^-)}{\Gamma (K^+ \to \pi^+ \pi^0 e^+ e^-) + \Gamma (K^- \to \pi^- \pi^0 e^+ e^-)}\sim \frac{e^2}{\Gamma_B}\bigg\langle|E_B||E_{DE}|\bigg\rangle\sin\delta\sin\Phi_{E},\label{eq:asymCP}
\end{align}
where we have defined $E_B={\cal{M}}(K^+\to \pi^+\pi^0\gamma^*)_{B}$ ({\it{cf.}} Eq.~(\ref{brem})). This interference already appeared in $K^\pm \to   \pi ^\pm \pi^0  \gamma$, with the lower bound $ A^{\pi^\pm\pi^0\gamma}_{CP}<1.5 \cdot 10^{-3}$ \cite{:2010uja}. We want to remark that, exactly as in $K^\pm \to   \pi ^\pm \pi^0  \gamma$, the appearance of the SM weak phases $\Phi_{E,M}$ occurs at ${\cal{O}}(p^6)$ in the chiral expansion of the direct emission form factors~\cite{Riazuddin:1993pn,Dib:1990gr}. Therefore, in MFV, $E_{DE}$ is not the one given in Eq.~(\ref{elmag}) but is related to the chromoelectric operator $Q_{7\gamma}$ discussed in Section~\ref{sec:V}. However, we note that ${\cal{O}}(p^4)$ contributions can be generated in non-MFV scenarios~\cite{Colangelo:1999kr,Tandean:2000qk}. 

The main advantage of $K^\pm\to\pi^\pm\pi^0e^+e^-$ is that one can also build a CP-violating observable not afflicted by strong phase suppression. This can be achieved through a charge asymmetry with the angular asymmetry discussed in Section~\ref{asymmLD}:  
\begin{align} 
A_{CP}^{\phi^*}&=\frac{\displaystyle\int_0^{2\pi}\frac{d\Gamma_{(K^+-K^-)}}{d\phi}d\phi^*}{\displaystyle\int_0^{2\pi}\frac{d\Gamma_{(K^++K^-)}}{d\phi}d\phi}\sim \frac{e^2}{\Gamma_B}\bigg\langle|E_B| |M_{DE}|\bigg\rangle_{\phi^*-asym}\cos\delta\sin\Phi_{M},\label{CPphi1}
\end{align}
where the magnetic amplitude $M_{DE}$ is, again, a chirally-suppressed ${\cal{O}}(p^6)$ contribution~\cite{Riazuddin:1993pn,Dib:1990gr}. Notice that this is a genuine CP-violating contribution of $K^\pm\to\pi ^\pm \pi^0 e^+e^-$, which comes from the fact that the virtual photon can generate different helicity amplitudes. We want to remark that this term might give a CP-violating signal 2 orders of magnitude larger than the one in $K^+ \to \pi^+ \pi^0 \gamma$, {\it{i.e.}} $A_{CP}^{\pi^\pm \pi^0\gamma}$: (i) it avoids strong phase suppression and (ii) a better detection can be achieved focussing on the central part of the Dalitz plot,\footnote{We are here assuming that the magnetic contribution induced by $Q_{7\gamma}$ does not differ kinematically from the ${\cal{O}}(p^4)$ one considered in Section~\ref{sec:III}.} where, for moderate values of $q^2$, there are strong enhancements of the signal over background ratio (see Table~4). 

The observables $A_{CP}$ and $A_{CP}^{\phi^*}$ considered before select CP violation in $\gamma$-mediated $K^\pm \to   \pi ^\pm \pi^0 e^+e^-$. One can also access CP violation in direct short distance operators to $K^\pm \to   \pi ^\pm \pi^0 e^+e^-$ by defining an alternative $\phi$-asymmetry, namely
\begin{equation}
\int_0^{2\pi}d{\tilde{\phi}}\equiv \left[\int_0^{\pi/2}+\int_{\pi/2}^{\pi}-\int_{\pi}^{3\pi/2}-\int_{3\pi/2}^{2\pi}\right]d\phi,
\end{equation}
which naturally leads to the observable\footnote{The terms proportional to $G_{13}$ and $G_{23}$ can in principle be accessed with a different angular asymmetry, but they are suppressed by strong phases.} 
\begin{align} 
A_{CP}^{\tilde{\phi}}&=\frac{\displaystyle\int_0^{2\pi}\frac{d\Gamma_{(K^+-K^-)}}{d\phi}d{\tilde{\phi}}}{\displaystyle\int_0^{2\pi}\frac{d\Gamma_{(K^++K^-)}}{d\phi}d\phi}\sim \frac{e}{\Gamma_Bq^2}\bigg\langle G_{12}|{\cal{F}}_1| |{{F}}_{B}|\bigg\rangle_{\tilde{\phi}-asym}\cos\delta\sin\Phi.\label{CPphi2}
\end{align}
Notice that $A_{CP}^{\tilde{\phi}}$ actually selects the CP-violating piece of $A_P^{(S)}$ defined in the previous Section. There it was competing with CP-conserving terms and, since it was a subdominant effect, it was neglected in Eq.~(\ref{eq:EMinterf}).

Due to the structure of the NA62 detector, only the forward direction in $\phi$ can be accessed. This suggests to define the forward asymmetry:
\begin{align}
A_{CP}^{(f)}=\frac{\displaystyle\int_0^{\pi/2}\frac{d\Gamma_{(K^+-K^-)}}{d\phi}d\phi-\int_{3\pi/2}^{2\pi}\frac{d\Gamma_{(K^+-K^-)}}{d\phi}d\phi}{\displaystyle\int_0^{\pi/2}\frac{d\Gamma_{(K^++K^-)}}{d\phi}d\phi+\int_{3\pi/2}^{2\pi}\frac{d\Gamma_{(K^++K^-)}}{d\phi}d\phi}.
\end{align}
Notice that, in view of the observables defined in Eqs.~(\ref{CPphi1}) and (\ref{CPphi2}), they both contribute. Therefore, this forward asymmetry is a way to optimize the detection of direct CP-violating phases.
 
It is instructive to compare the situation with $K_L\to\pi^+\pi^-e^+e^-$. There CP observables are dominated by indirect CP violation, which overshadows any direct CP contribution. One way to isolate direct CP violation was put forward in Ref.~\cite{Elwood:1995xv}, where the analogue of $A_P^{(S)}$ was considered. The advantage of $K^+\to\pi^+\pi^0e^+e^-$ is that direct CP violation is free from indirect CP violation backgrounds. This, added to the fact that NA62 will collect $K^+$ decays with large statistics, makes the study of this channel appealing.   


\section{Conclusions}\label{sec:VII}

In this work, we have studied different aspects of the decays $K^{\pm}\to\pi^{\pm}\pi^0e^+e^-$. At long distances, we have performed a thorough Dalitz plot study. In the past, cuts in the leptonic invariant mass $q^2$ were shown to be useful to extract information about the different contributions of this decay. Here, we have shown that additional cuts on the variables $(E_{\gamma}^*,T_c^*)$ can provide even cleaner signals for a precise extraction of all contributions by concentrating on different corners of the phase space.

In particular, interesting information about the direct emission (electric and magnetic) contributions can be extracted through their inteference with the dominant Bremsstrahlung contribution. The electric interference was already determined from $K^+\to\pi^+\pi^0\gamma$ by the NA48/2 collaboration, leading to a discrepancy in sign with the theoretical estimates. $K^+\to\pi^+\pi^0e^+e^-$ offers an alternative measurement with the potential to sort out the discrepancy. The magnetic interference term is however a genuine feature of $K^+\to\pi^+\pi^0e^+e^-$ that will allow for a measurement of the sign of the magnetic term, something relevant for chiral tests of the anomalous weak chiral Lagrangian. Additionally, it provides a clean determination of the FSI phases in $K^{\pm}\to\pi^{\pm}\pi^0$. In that sense, $K^+\to\pi^+\pi^0e^+e^-$ can be considered a laboratory for chiral tests.  

Short distances within and beyond the Standard Model can be probed in this decay channel without resorting to charge asymmetries: observables with P violation in the dilepton pair coupling are already a signature of short-distance physics. Thus, even at this stage of the NA62 experiment (with only $K^+$), short-distance analyses can be performed. Regarding the charge asymmetry, an interesting feature of $K^{\pm}\to \pi^{\pm}\pi^0e^+e^-$ over its neutral counterpart $K_L\to\pi^+\pi^-e^+e^-$ is that short-distance observables are not contaminated by indirect CP violation. We have also explored the potential of charge asymmetries to uncover new physics, should NA62 eventually start data recollection on $K^-$. With the help of the Dalitz plots and assuming Minimal Flavor Violation, a very conservative BSM scenario, one can show which are the optimal regions to look for signals of CP violation.  


\section*{Acknowledgments}
We are very grateful to Augusto Ceccucci, Sergio Giudici, Mauro Raggi and Marco Sozzi for useful discussions. O.~C. wants to thank the University of Naples and G.~D'A the University of Val\`encia and IFIC for very pleasant stays during the different stages of this work. L.~C.~ and G.~D'A. are supported in part by MIUR, Italy, under project 2005-023102 and by Fondo Dipartimentale per la Ricerca 2009. O.~C. is supported by MICINN (Spain) under Grants FPA2007-60323, by the Spanish Consolider Ingenio 2010 Programme CPAN (CSD2007-00042) and by the DFG cluster of excellence 'Origin and Structure of the Universe'. D.~N.~G. is supported in part by the NSF of China under Grant No. 11075149 and the 973 project under Grant No. 2009CB825200.

\appendix
\section{Appendix}

For completeness we here list the differential decay rate for the Bremsstrahlung contribution of Eq.~(\ref{results}) with isospin breaking corrections taken into account:
\begin{align}
\frac{d^3\Gamma_{B}}{dE_{\gamma}^*dT_c^*dq^2}&=\frac{\alpha^2|{\cal{M}}_K|^2(2m_l^2+q^2)}{48\pi^3 m_K^3q^4(q^2-2E_{\gamma}^*m_K)^2}\sqrt{1-\frac{4m_l^2}{q^2}}\frac{q^6+\lambda_1^{\chi}q^4+\lambda_2^{\chi}q^2+\lambda_0^{\chi}}{{[(m_K-2(m_{\pi^+}+T_c^*+E_{\gamma}^*))-\chi(m_K-2m_{\pi})]^2}},\nonumber\\
\end{align}
where we have defined
\begin{align}
\lambda_1^{\chi}&=\lambda_1-2m_K^2\chi-m_K^2\chi^2,\nonumber\\
\lambda_2^{\chi}&=\lambda_2+2m_K^2\chi\left(m_K^2-2m_K (2m_{\pi}+3 T_c^*)+4 m_{\pi} (m_{\pi}+T_c^*)\right),\nonumber\\
&+m_K^2\chi^2 \left(4E_{\gamma}^*m_K-3m_K^2+4m_Km_{\pi}+4m_{\pi}^2\right),\nonumber\\
\lambda_0^{\chi}&=-4m_K^2\sigma -8 m_K^3\lambda_3\chi(E_{\gamma}^*-m_K+2
   (m_{\pi}+T_c^*))-4m_K^4\chi^2(E_{\gamma}^*-m_K+2m_{\pi})^2,
\end{align}
and the isospin-breaking parameter
\begin{align}
\chi=\frac{m_{\pi^+}-m_{\pi^0}}{m_K}.
\end{align}

\end{document}